\documentclass[12pt]{article}
\usepackage{psfig}
\begin {document}
\begin{flushright}
{\small
SLAC--PUB--8965 \\
October 2001}
\end{flushright}

\begin{center}
{{\bf\LARGE \boldmath $B$ Physics and  $CP$ \unboldmath
Violation}\footnote{Work supported by Department of Energy
contract DE--AC03--76SF00515.}}

\bigskip
{\it Helen Quinn \\
Stanford Linear Accelerator Center \\
Stanford University, Stanford, California 94309 \\
E-mail:  quinn@slac.stanford.edu}
\medskip
\end{center}

\vfill

\begin{center}
{\bf\large Abstract }
\end{center}

These lectures provide a basic overview of topics related to the
study of $CP$ Violation in $B$ decays.  In the first lecture, I
review the basics of discrete symmetries in field theories, the
quantum mechanics of neutral but flavor-non-trivial mesons, and
the classification of three types of $CP$ violation \cite{texts}.
The actual second lecture which I gave will be separately
published as it is my Dirac award lecture and is focussed on the
separate topic of strong $CP$ Violation.  In Lecture 2 here, I
cover the Standard Model predictions for neutral $B$ decays, and
in particular discuss some channels of interest for $CP$ Violation
studies. Lecture 3 reviews
 the various tools and techniques used to deal with the
hadronic physics effects. In Lecture 4, I briefly review the
present and planned experiments that can study $B$ decays.  I
cannot teach all the details of this subject in this short course,
so my approach is instead to try to give students a grasp of the
relevant concepts and an overview of the available tools. The
level of these lectures is introductory.
 I will provide some references to  more detailed treatments
and current literature, but this is not a review article so I do
not attempt to give complete references to all related literature.
By now there are some excellent textbooks that cover this subject
in great detail \cite{texts}. I refer students to these for more
details and for more complete references to the original
literature.

\vfill

\begin{center}
{\it Lectures given at the \\
Particle Physics School \\
International Centre for theoretical Physics\\
Trieste, Italy\\
2--6 July 2001 }\\
\end{center}

\vfill \newpage

\setcounter{footnote}{0}

\renewcommand{\bar}[1]{\overline{#1}}

\section{Lecture 1: Preliminaries: Symmetries, Hermiticity, Rephasing Invariance}

We begin with the basics of symmetries in Lagrangian Field Theory.
Physicists use the term symmetry to denote an invariance of the
Lagrangian, and thus of the associated equations of motion, under
some change of variables.  Such changes can be local, that is
coordinate dependent, or global; and they can be a continuous set
or a discrete set of changes.  The value of such symmetries lies
in the simplification they achieve by limiting possible terms in
the Lagrangian and by their relationship to conservation laws and
the conserved quantum numbers that then characterize physical
states.  The invariance may be with respect to coordinate
redefinitions, as in the case of Lorentz Invariance, or field
redefinitions, as in the case of gauge invariance.  The particular
invariances of interest to us in these lectures are the global
discrete invariances known as $C$, $P$, and $T$. These are charge
conjugation or $C$ (replacement of a field by its
particle-antiparticle conjugate), parity or $P$ (sign reversal of
all spatial coordinates), and time reversal or $T$ (sign reversal
of the time coordinate, which reverses the role of in and out
states).  Table \ref{tab:1} shows the effect of these operations
on a Dirac spinor field $\psi$, and Table \ref{tab:2} summarizes
the effect of the particular combination $CP$ on some quantities
that appear in a gauge theory Lagrangian. In Table  \ref{tab:2},
the symbol $(-1)^\mu$ denotes a factor +1 for $\mu=0$ and -1 for
$\mu=1,2,3$.

\begin{table}[h]\caption{The operation of $P$,$C$, and $T$ on a Dirac spinor field}
\begin{center}
\begin{tabular}{rl}
\noalign{\smallskip}
$P\psi(t,x)P\!$& $=\gamma^0\psi(t,-x)$ , \\[1ex]
$T\psi(t,x)T\!$& $=-\gamma^1\gamma^3\psi(-t,x)$ , \\[1ex]
$C\psi(t,x)C\!$& $=-i(\bar\psi(t,x)\gamma^0\gamma^2)^T $
\end{tabular}
\end{center}
\label{tab:1}
\end{table}

\begin{table}[h]\caption{The effect of a $CP$ transformation on various quantities}
\begin{center}
\begin{tabular}{ccccc}
\noalign{\smallskip} term & $\bar\psi_i\psi_j$ &
$i\bar\psi_i\gamma^5\psi_j$ & $\bar\psi_i\gamma^\mu\psi_j$ &
$\bar\psi_i\gamma^\mu\gamma^5\psi_j$
\\[1ex]
$CP$-transformed term & $\bar\psi_j\psi_i $ &
$-i\bar\psi_j\gamma^5\psi_i$ &
$-(-1)^\mu\bar\psi_j\gamma^\mu\psi_i$ &
$-(-1)^\mu\bar\psi_j\gamma^\mu\gamma^5\psi_i$
\\[1ex]
term & $H$ & $A$ & $W^{\pm\mu}$ & $\partial_\mu$
\\[1ex]
$CP$-transformed term & $H$ & $-A$ & $-(-1)^\mu W^{\mp\mu}$
&$(-1)^\mu\partial_\mu$
\end{tabular}
\end{center}
\label{tab:2}
\end{table}

When constructing a field theory we always require locality, the
symmetries of Lorentz Invariance, and hermiticity of $\cal L$.
That is sufficient to make any field theory automatically also
invariant under the product of operations $CPT$.  In many
theories, for example for QED with fermion masses included, the
combination $CP$, and thus also $T$ are also separately automatic.
This is the reason why the experimental discovery that $CP$ is not
an exact symmetry of nature caused such a stir.  All the field
theories that had been studied up to that time had automatic $CP$
conservation.  So we need to examine how $CP$ non-conservation
manifests itself, and then ask what theories will give such
effects.

$CP$ non-conservation shows up, for example, as a rate difference
between two processes that are the $CP$ conjugates of one-another.
How can such a rate difference appear?  Consider a particle decay
for which two different terms in the Lagrangian (two different
Feynman diagrams) give possible contributions.  The amplitude for
such a process can be written as
\begin{equation}
A= A(A \rightarrow B) = g_1 r_1 e^{i\phi_1} + g_2 r_2 e^{i\phi_2}
\ .
\end{equation}
Here $g_1$ and $g_2$ are two different, possibly complex, coupling
constants in the theory.  The transition amplitudes corresponding
to each coupling are written as $re^{i\phi}$ to emphasize that
they too can have both a real part or magnitude and a phase or
absorptive part.  The physical source of this phase is that there
may be multiple real intermediate states which can contribute to
the process in question via rescattering effects.  In the jargon
of the field the phases $\phi$ are called strong phases because
the rescattering effects among the various coupled channels are
dominated by strong interactions. These phases are the same for a
process and its $CP$ conjugate because the $CP$-related sets of
intermediate states must contribute the same absorptive part to
the two processes.  The phases of the coupling constants are often
called weak phases because, in the Standard Model, the relevant
complex couplings are in the weak interaction sector of the
theory.  When we look at the amplitude for the $CP$ conjugate
process we find
\begin{equation}
\bar A = A(\bar A\rightarrow \bar B) = g_1^* r_1 e^{i\phi_1} +
g_2^* r_2 e^{i\phi_2} \ .
\end{equation}
Note that the phases of the coupling constants change sign between
any process and its $CP$ conjugate process, while the strong
phases, which arise from absorptive parts in the amplitudes, do
not.

So now let us calculate the $CP$-violating difference in rates for
these two processes. With a little algebra we find
\begin{equation}
|A|^2 -|\bar A|^2 = 2r_1 r_2 {\rm Im} g_1 g_2^* \sin (\phi_1
-\phi_2) \ . \label{ratediff}
\end{equation}
This shows that the effect will vanish if the two coupling
constants can be made relatively real. In addition it depends on
the difference of strong phases in the two amplitude
contributions, and vanishes if this quantity is zero.  Such a $CP$
violation in the comparison of two $CP$-related decay rates is
often called direct $CP$ violation.  I prefer the more descriptive
term $CP$ violation in the decay amplitudes. Whatever you choose
to call it, this effect is characterized by the condition $|\bar
A/A|\neq 1$.   It is obvious that in any process where there is
only a single contributing term in the decay amplitude the phase
of the coupling constant is irrelevant and $|\bar A/A|=1$ is
automatic.
 You need two different couplings contributing, with
non-zero relative phase of the two couplings to see any $CP$
violation.

This statement applies for al types of $CP$ violation. The phase
of any single complex coupling in a Lagrangian is not a physically
meaningful quantity. In general it can be redefined, and even made
to vanish by simply redefining some field or set of fields by
appropriate phase factors.  But such rephasing of fields can never
change the relative phase between two couplings (or products of
couplings) that contribute to the same process.  Both contributing
terms must involve the same nett set of fields, and hence both
change in the same way under any rephasings of those fields. These
rephasing-invariant quantities are the physically meaningful
phases in any Lagrangian, the existence of such a quantity signals
the possibility of $CP$ violation.

The second feature we note is that the $CP$-violating rate
difference in Eq.~(\ref{ratediff}) also depends on a difference of
strong phases. Typically, this makes it difficult to calculate.
Strong phases are, in general, long-range strong interaction
physics effects, not amenable to perturbative calculation. One of
the things that makes the decays of neutral but flavored mesons
particularly interesting is that there we find other types of
$CP$-violation effects where the role played here by the strong
phases is replaced by other coupling constant phases, those
relevant to the processes that mix the meson with its $CP$ (and
thus also flavor) conjugate meson. In such a case we may be able
to relate a measured $CP$ violation directly to phase-differences
in the Lagrangian couplings, with no need to calculate any
strong-interaction quantities. Only in the case of neutral but
flavor non-trivial mesons can such mixing-dependent effects occur.

We have seen that only a theory with two coupling constants that
are not relatively real can give $CP$ violation.  Thus we only can
have $CP$ violation in a theory where there is some set of
couplings for which rephasing of all fields cannot remove all
phases. $CP$ conservation is automatic for any theory for which
the most general form of the Lagrangian allows all complex phases
to be removed by rephasing of some set of fields.  Let us examine
a few of the terms that occur in the QED Lagrangian to see why
$CP$ conservation is automatic in that theory.  For the gauge
coupling terms we have, after requiring hermiticity
\begin{equation}
g A^\mu \bar \psi \gamma_\mu \psi +g^* A_\mu \bar \psi \gamma^\mu
\psi \ .
\end{equation}
Thus hermiticity clearly makes the QED gauge coupling real, ($g
+g^*$), because the term it multiplies is itself a hermitian
quantity. After imposing hermiticity you will find that the
fermion mass term must take the form
\begin{equation}
Re( m)\bar \psi \psi +i{\rm Im} (m)\bar \psi \gamma_5\psi
\end{equation}
for any complex $m$. Hermiticity alone does not require that the
fermion mass be real, but it does require that the imaginary part
multiplies a factor of $\gamma_5$.
 But a chiral rephasing of the fermion
field $\psi \rightarrow e^{i\phi \gamma_5}\psi $ can be made. This
does not change the kinetic or gauge coupling terms at all. In
QED, one can always choose the angle $\phi$ in this rotation
 in such a way that it makes m a real quantity.  This tells us that, in such a
theory, the phase of m is not a physically meaningful quantity.
Hence the theory is indeed automatically $CP$ conserving for any
choice of $m$. (It is merely for convenience that we always choose
to write QED with real particle masses; it is unnecessary to
include additional parameters that you know are irrelevant to
complicate your calculations.) Tomorrow we will see that this same
rephasing is not so innocuous in $QCD$, and how this leads to the
strong $CP$ problem \cite{dirac}.

Given these examples you may be beginning to wonder how we ever
get a $CP$ violating coupling into a Lagrangian field theory. That
is the question that puzzled everyone in 1964.  The trick is to
have a sufficient number of different terms in the Lagrangian
involving the same set of fields.  For example imagine a theory
with multiple flavors of fermions and multiple scalar fields.  In
such a theory there can be Yukawa couplings of the form $Y_{ijk}
\phi_k\bar \psi_i \psi_j$. Hermiticity then requires only that we
also have a term $Y^*_{ijk} \phi^*_k \bar \psi_j\psi_i$ in the
Lagrangian.  Note that this is a different product of fields from
the original term, so hermiticity does not disallow phases for the
various $Y_{ijk}$ in such a theory.  But we still must ask whether
we can make every such coupling real, by systematically redefining
the phases of the various fields.  That depends on the details of
the theory.  As we add more fields of a given type, either
fermions or scalars, the number of possible coupling terms grows
more rapidly than total number of fields.  With enough fields of
the each type there will be more couplings that there are possible
phase redefinitions, and then not all couplings can be made real
by rephasing the fields.

We can always make all couplings real by imposing $CP$ invariance
as a postulate, but it no longer an automatic feature of the
theory.  The Standard Model with only one Higgs doublet and only
two fermion generations has automatic $CP$ invariance; all
possible couplings can be made simultaneously real (ignoring for
now the issue of strong $CP$-violation via a QCD-theta parameter).
Adding one more generation of fermions or adding an additional
Higgs doublet with no further symmetries imposed opens up the
possibility of $CP$ violating couplings \cite{CPVpossibilities}.
The three generation Standard Model with a single Higgs doublet
has only one $CP$-violating parameter, that is only one
independent phase difference survives after as many couplings as
possible are made real by field rephasing. This means that all
$CP$-violating effects in this theory are related. That is what
makes it so interesting to test the pattern of $CP$ violation in
$B$ decays. Here there are many different channels in which
possible $CP$-violating effects may be observed. In the Standard
Model there are predicted relationships between these effects, and
between $CP$ violating effects and the values of other
$CP$-conserving Standard Model parameters. Thus the patterns of
the $B$ decays, as well as their relationships to the observed
$CP$ violation in $K$-decays, provide ways to test for the effects
of physics beyond the Standard Model. Such effects can disrupt the
predicted Standard Model relationships between the  different
measurements.

\subsection{Quantum Mechanics of Neutral Mesons}

We now we turn to a general discussion of the physics of flavored
neutral mesons, those made from different quark and antiquark
types of the same charge. These are the $K$, $D$, $B_d$ and $B_s$
mesons, which we denote generically by $M^0$.  (I use the notation
$B_d$ as a reminder of the quark content, even though the official
name of this particle is simply $B^0$.)  There is a beautiful
quantum mechanical story here.  In each case there are two
$CP$-conjugate flavor eigenstates, $M^0 = \bar q q^\prime$ and
$\bar M^0 =\bar q^\prime q$.  In general $CP M^0 = e^{i\xi} \bar
M^0 $. The phase $\xi$ is convention dependent and can be altered
by redefining one or other of the quark fields by a phase.  In
much of the literature on this subject the convention $\xi=0$ is
chosen without comment, but elsewhere $\xi=\pi$ is used.  Physical
results are convention independent, but only as long as you
consistently use the same convention.  You can get into trouble if
you combine formulae taken from two different sources without
first checking that both are using the same convention.  From this
point on I will use the convention $\xi=0$; if you want to see the
equations with arbitrary phase factors explicitly displayed, go to
the textbooks \cite{texts}.

Let us for the moment assume that $CP$ is a symmetry of our
theory. What does this tell us about the neutral mesons? It says
that the physical propagation-eigenstates of the system, that is
the particles which propagate with a distinct mass (and lifetime),
must be eigenstates of $CP$.  These are the combinations $(M^0 \pm
\bar M^0)/ \sqrt{2}$. Particles produced by the strong
interactions are produced as flavor eigenstates.  This means
initially one always has a coherent superposition of the two $CP$
eigenstates.  Then as time goes on, because of the difference in
masses of these two states, their relative phases change. Thus, if
both states are long-lived enough, the flavor composition
oscillates.
 However there is also a difference in lifetime of the two $CP$ eigenstates.
If this is large then eventually the shorter-lived eigenstate
decays away. Once one of the two mass eigenstates has decayed the
other combination dominates, terminating the flavor oscillation
and giving essentially a fixed admixture from that time on (in
vacuum).  For the kaon system the difference in lifetime is large
compared to the difference in mass, so one does not talk about
kaon oscillation, but rather about long-lived and short-lived
states.  Conversely for $B_d$ the mass difference is large
compared to the width difference, and one can discuss either
oscillating flavor states, or, discuss the same phenomena in the
language of mass eigenstates, $B_{H={\rm heavy}}$ and $B_{L={\rm
light}}$.  For the $B_s$ both the mass and lifetime differences
must be both be considered in analyzing the evolution of states.
For the $D$ mesons, in contrast, the mass and width differences
are both small
 in the Standard model. Thus both mass eigenstates decay before any significant
oscillation occurs. These particles are thus typically described
in terms of flavor eigenstates.  Experimental searches for
evidence of mixing (mass or width differences) for the $D^0$
states are another way to seek non-Standard Model physics effects,
since the effect as predicted in the standard Model is small
\cite{dmixing}.

Notice that the peculiar phenomenon of oscillating particles, here
and in the neutrino case as well, occurs only if you insist on
describing the process in terms of flavor eigenstates.  The more
physical description is to use the mass eigenstates as the things
you call particles (as we do for the quarks themselves).  Then all
that changes with time is the proportion of the two eigenstates
that are present, because of their different half-lives, and the
relative phase of the two states, because of their different
masses.

Now let us review the story of $CP$ for neutral $K$ mesons.  The
flavor quantum number strangeness is conserved in strong
interactions. Strangeness-changing weak decays are suppressed by
the Cabibbo factors tan($\theta_{\rm Cabibbo}$) compared to
strangeness conserving $u <-> d$ transitions.  This first fact
means strange mesons are typically pair produced, the second that
they are relatively long lived.  The assumption of
$CP$-conservation in neutral Kaon decays ``explains'' the
observation of the two very different half-lives for neutral
kaons.  If $CP$ were exact, then only the $CP$-even state, $K_{\rm
even} =(K^0 +\bar K^0)/\sqrt{2}$, can decay to two pions, since a
spin zero neutral state of two pions can only be $CP$-even.  (By
Bose statistics, it can have no I=1 part.) Three-pion final states
can be either $CP$-even or $CP$-odd. But the phase space for the
three pion decay of a neutral kaon is quite small compared to that
for two pions. This predicts two very different half-lives for the
two $CP$-eigenstates.  They are different, in fact, by more than a
factor of ten.

This successful picture was challenged in 1964 by the discovery by
Christensen, Cronin, Fitch and Turlay \cite{ccft}, that the
long-lived (and hence putatively $CP$-odd) kaon state did indeed
sometimes decay into the $CP$-even two pion state.  This result
immediately shows that $CP$-invariance  is violated. Comparison of
the rates for charged and neutral pions further showed that the
violation is principally in the fact that the mass eigenstate does
not have a unique $CP$. This result was initially very puzzling.
Until then almost any field theory that had been considered as a
realistic physical theory had automatic $CP$ conservation once the
other desired symmetries of were imposed.  Now, however, we know
that the three generation Standard Model in its most general form
includes one $CP$-violating parameter in the matrix of weak
couplings, which is called the CKM matrix (for Cabibbo, Kobayashi
and Maskawa). Thus $CP$ violation {\em per se} is no longer a
puzzle, but rather a natural part of the Standard Model.  What we
do not yet know is whether the Standard Model correctly describes
the $CP$-violation found in nature.  Exploration of that question
is a major goal of the B-physics program.

Any theory for physics beyond the Standard Model will have, in
general, possible additional $CP$-violating parameters.  Any
further fields, such as any additional Higgs fields, can introduce
further $CP$-violating couplings. Such effects may then enter into
$B$ decay physics. For example, in many models additional Higgs
particles lead to additional contributions to $B^0$-$\bar B^0$
mixing.  This in turn gives possible deviations from the patterns
predicted by the Standard Model for $CP$-violation in $B$ decays.
One of the motivations to search for such effects is that it is
not possible to fit the observed matter-antimatter imbalance (or
rather the consequent matter to radiation balance) of the Universe
with the $CP$-violation in the quark mixing matrix as the only
such effect \cite{baryofailure}. (This failure suggests that there
must be additional sources
 of $CP$-violation beyond those in the quark coupling matrix of the Standard Model,
but does not require that any such effects will be apparent in $B$
decays.)

Even with no other new particles, an extension of the Standard
Model to include neutrino masses now appears to be needed.  Then
the weak couplings of the neutrino mass eigenstates are given by a
CKM-like matrix.  This introduces the possibility of further
$CP$-violating parameters.  Indeed if the neutrinos have Majorana
type masses there are more $CP$-violating parameters in this
matrix than in the quark case  \cite{numass}. These parameters
will be very difficult to determine and they play essentially no
role in $B$ physics.  However they may have played an important
role in the early universe, giving the matter-antimatter imbalance
via leptogensis \cite{leptogenesis}. I will not discuss neurtrino
masses further in these lectures.

As I will discuss tomorrow \cite{dirac}, once there is any $CP$
violation in the Standard Model theory it becomes a problem to
understand how it happens that $CP$ is conserved in the strong
interaction sector of the theory.  Experiment tells us this is so
to very high accuracy, chiefly via the upper limit on the electric
dipole moment of the neutron.  This result tells us that, far as
the $CP$-violating effects that we want to explore in $B$ decays
go, we can ignore strong $CP$ violation.  So apart from tomorrow's
Dirac lecture, I will not discuss it further in this series of
talks.

\boldmath
\subsection{General Formalism for Neutral Mesons with $CP$ Violation}
\unboldmath

Once we know that $CP$ is not a symmetry of our theory we must
allow a more general form for the two mass eigenstates of neutral
but flavored mesons.  In the following I use the convention that
these two states are defined to be $M_H$ and $M_L$ where the $H$
and $L$ stand for heavy and light, which really means heavier and
less heavy, since the mass difference may indeed be quite tiny.

I define the two eigenstates to be
\begin{equation}
M_{H } = p M^0 +  q \bar M^0 \qquad M_{L} = p M^0 -  q \bar M^0 ,
 \label{masseigen}
\end{equation}
where $|p|^2 + |q|^2 = 1$. Note that this equation is again
convention dependent, I have not specified a sign or phase for
$q$, but I have defined the more massive state to be the one with
a plus sign before $q$. In combination with my convention that $CP
M^0 = \bar M^0$ this makes the phase of $q$ a meaningful quantity.
(Be aware however that, once again, other conventions are also
used in the literature.)

\begin{figure}[htb]
\centerline{\hbox{ \psfig{figure=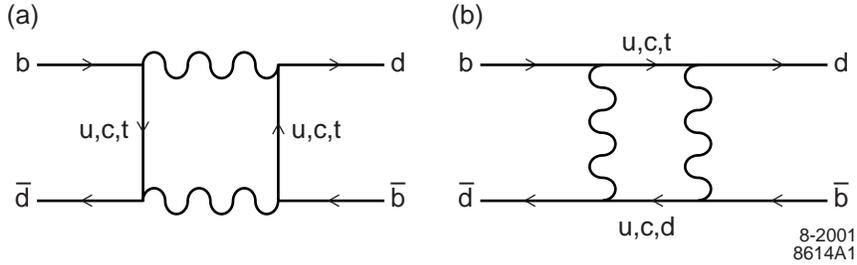}}}
 \caption[*]{\footnotesize Leading Diagrams for $B\bar B$ Mixing
 in the Standard Model} \label{fig:1}
\end{figure}

The quantity q/p is determined from the mass and mixing matrix for
the two-meson system, ${\cal M} = M +i \Gamma$.  This matrix is
written in the basis of the two flavor eigenstates.  Note that
both M and $\Gamma$ are complex 2$\times$2 matrices, $M$ is
hermitian and $\Gamma$ is anti-hermitian.  The off-diagonal (or
mixing) elements are calculated from Feynman Diagrams that can
convert one flavor eigenstate to the other.  In the Standard Model
these are dominated by the one loop box diagrams, shown in Fig.
\ref{fig:1}. Actual calculation of such quantities will be
discussed in later lectures, for now we simply note that they
exist. Then
\begin{equation}
q/p = \frac{\Delta M -i/2\Delta\Gamma}{2(M_{12} -i/2 \Gamma_{12})}
= \frac{M_{12} -i/2 \Gamma_{12}}{2(\Delta M -i/2\Delta\Gamma)}\ .
\label{qoverp}
\end{equation}
Notice that the two mass eigenstates of this mixed system do not
have to be orthogonal, in fact in general they will not be so,
unless $|q/p|=1$.

\boldmath
\subsection{The Three Types of $CP$ Violation}
\unboldmath

In the above discussion we have already mentioned two possible
ways that $CP$ violation can occur.  The first was $CP$ violation
in the decay, or direct $CP$ violation, which requires that two
$CP$-conjugate processes to have differing absolute values for
their amplitudes.  A second possibility, seen for example in $K$
decays, occurs if $|q/p| \neq 1$.  It is very clear in this case
that no choice of phase conventions can make the two mass
eigenstates be $CP$ eigenstates. This is generally called
$CP$-violation in the mixing.  As we will see later, in decays of
the neutral mesons to a $CP$-eigenstate $f$, there is a third
possibility. This can occur even when both the ratio of amplitudes
and the quantity $q/p$ have absolute value 1. The $CP$ violation
effects in such decays will be shown to depend only on the
deviations from unity of the parameter
 $\lambda_f =(q/p) A(\bar B^0 \rightarrow  f)/A(B^0 \rightarrow f)$ .
 The third option is
$CP$ violation in the interference between  decays to $f$ with and
without mixing.  This effect is proportional to the imaginary part
of $\lambda_f$ and thus can be non-zero even when the absolute
value satisfies $|\lambda_f|=1$.  Decays where this latter
condition is true are particularly interesting. In such cases one
can interpret any observed asymmetry as a direct measurement of
some difference of phases of CKM matrix elements, with no
theoretical uncertainties.  We will see this in more detail in the
next lecture.

\boldmath
\section{Lecture 2: Standard Model Predictions for \hfill\break
$CP$ Violations in $B$ Decays} \unboldmath

\subsection{CKM Unitarity}

The CKM matrix of quark weak couplings has been discussed in some
detail in previous lecture series in this school. It can be
written, in the Wolfenstein parameterization
\cite{wolfensteinparam}, as
\begin{eqnarray}
 V &=&\pmatrix{V_{ud}&V_{us}&V_{ub}\cr
          V_{cd}&V_{cs}&V_{cb}\cr
          V_{td}&V_{ts}&V_{tb} }\nonumber \\[1ex]
&\simeq& \pmatrix{1-\lambda^2/2&\lambda &A\lambda^3(\rho-i\eta)\cr
                 -\lambda    &1-\lambda^2/2&A\lambda^2\cr
     A\lambda^3(1-\rho-i\eta)&-A\lambda^2&1}
     +  O(\lambda^4)~.
\label{ckm}
\end{eqnarray}

In the previous lecture I talked about the ability to remove, or
move, a complex phase of a coupling by redefining the phase of any
field involved.  This parameterization corresponds to a particular
choice of phase convention which eliminates as many phases as
possible and puts the one remaining, possibly large, complex phase
in the matrix elements $V_{ub}$ and $V_{td}$.

In this convention the upper right off-diagonal elements define
the parameters.  The parameterization is a convenient way to make
the unitarity of the matrix explicit, up to higher order
corrections in powers of $\lambda \equiv V_{us}$.  (The higher
order terms may also have phases, as required by the unitarity
relationships, but bring in no new independent phase parameters.)
The quantity $\lambda$ is essentially the sine of the Cabibbo
angle.  It is a small number, of order 0.2.  Wolfenstein's
parameterization uses powers of $\lambda$ is a convenient way to
keep track of the relative sizes of the terms in the matrix.  The
other independent magnitude parameters $A$ and $\rho^2 +\eta^2$
are known to be roughly of order unity.  There is no theory behind
which powers of $\lambda$ enter each term.  The Wolfenstein
parameterization simply summarizes the observations in a neat way.
The fact that $V_{cb}$ and $V_{ub}$ are both small (of order
$\lambda^2$ and $\lambda^3$ respectively in Wolfenstein's
parameterization) is responsible for the relatively long lifetimes
of $B$-mesons (and $b$-containing baryons too). This is a
fortunate property; it is essential to the feasibility of most
$B$-physics experiments because it allows us to identify $B$
decays by the spatial separation of the decay vertex from the
production point. It is an observational fact, not a theoretical
prediction.

Independent of the parameterization used, in the three generation
Standard Model the CKM matrix must be unitary.  This leads to a
number of relationships among its elements of the form
[(row)*x(column)]=0. Examples are
\begin{eqnarray}
V_{ud}V^*_{us} +V_{cd}V^*_{cs}+ V_{td}V^*_{ts} &=&0 \hspace{.5in} {a}\nonumber\\
V_{us}V^*_{ub} +V_{cs}V^*_{cb}+ V_{ts}V^*_{tb} &=&0\hspace{.5in} {b}\\
V_{ub}V^*_{ud} +V_{cb}V^*_{cd}+ V_{tb}V^*_{td} &=&0\hspace{.5in}
{c} \ .\nonumber \label{unitarity}
\end{eqnarray}
In the Wolfenstein parameterization the relationship that arises
from unitarity can be used to express the diagonal and lower left
hand elements of the matrix in terms of the upper right elements,
to any desired order in $\lambda$.  The form given above drops
terms of order $\lambda^4$ and above.

It is a trivial fact that any relationship of the form of a sum of
three complex numbers equal to zero can be drawn as a closed
triangle in the complex plane.  Hence these, and the other similar
relationships, are referred to as the Unitarity Triangle
relationships.  The fact that there is only one independent
$CP$-violating quantity in the CKM matrix can be expressed in
phase-convention-invariant form by defining the quantity $J$,
called the Jarlskog invariant for Cecilia Jarlskog who first
pointed out this form \cite{jarlskog},
\begin{equation}
{\rm Im} V_{ij}V_{kl}V_{il}^*V_{kj}^* = J \Sigma_{m,n =1}^3
\epsilon_{ikm}\epsilon_{jln} \label{Jarlskog}
\end{equation}
where $i,j,k,l$ run over the values $1,2,3$ and $\epsilon_{ijk} $
takes the value +1 if the three indices are all different and in
cyclic order, and -1 if they are all different and in anti-cyclic
order, but is zero if any two are the same.  All the unitarity
triangles have the same area, $J/2$. This area shrinks to zero if
the $CP$-violating phase differences in the matrix vanish.

Notice however that, while the triangles have the same area, the
three examples given above are triangles of very different shapes.
Triangle ${a}$ has two sides of order $\lambda$ and one of order
$\lambda^5$. It would be very difficult to measure the area using
such a triangle.  Triangle ${b}$ is a little better, but still a
has one small angle, its larger sides are of order $\lambda^2$
while its small side is of order $\lambda^4$ giving an angle of
order $\lambda^2$.  Finally triangle ${c}$ is the most
interesting, because it has all three sides of order $\lambda^3$
so all three angles are {\em a priori} of comparable and large
magnitude. The price one pays is that all the sides are small, but
this is not as serious as the problem of measuring an asymmetry
proportional to a very small angle. This triangle is the one most
often discussed in relation to $B$-meson decays.  Since these
angles are large one expects some channels in both $B_d$ and $B_s$
decays with order 1 $CP$-violating asymmetries .

\subsection{Fixing the Parameters}

The triangle is conventionally drawn by dividing all sides by
$V_{cb}V_{cd}^*$, which gives a triangle with base of unit length
whose apex is the point ($\rho,\eta$) in the complex plane. Prior
to considering the asymmetry measurements we can try to determine
the shape of this triangle from measurements of $CP$-conserving
quantities which fix the sides, plus the measured $CP$ violation
in $K$-decays. Notice that this information is already sufficient
(in principle) to over constrain the set of parameters.

The quantity $V_{cb}$ is determined from $B$ decays to charmed
final states, $V_{ub}$ from final states with no charm, while
measurements of the $B_d$ and $B_s$ mass differences constrain
$V_{td}$.  The $CP$ violation in $K \rightarrow \pi \pi$ gives an
allowed band for the apex of the triangle. In each case there is
both an experimental uncertainty in the measurement and a
theoretical uncertainty in the relationship between the measured
quantity and the theoretical parameter(s). The theoretical
uncertainties dominate.  They are typically not statistical in
nature, but rather have to do with the part of the calculation
which involves models or approximations needed to allow for strong
interaction physics effects.  There is a large literature by now
on the topic of how best to combine the various measurement and
deal with both statistical and theoretical uncertainties
\cite{traingleconstraints}.

New measurements from Belle and BaBar on a $CP$ asymmetry in
$B$-decays constraining the angle at the lower left of the
triangle have recently been announced \cite{newtwobeta}.  This is
one measurement where the theoretical uncertainties are very
small, so the constraint will improve as the statistics of the
measurement improve for some time to come. So far all the various
results give a consistent picture; the Standard Model fits the
data. This means that, within the ranges of the various
theoretical uncertainties, there is a region of possible choices
for the Lagrangian parameters that are consistent with all data.

One hope of many physicists involved in the large effort in $B$
physics is that at some point some measurements will give
discrepant answers for some Standard Model parameters or
predictions. This would be evidence for physics beyond the
Standard Model, and cause for much excitement in the physics
community. If results for some set of measurements should begin to
look discrepant, then the question of the statistical significance
of the discrepancy will be much debated, as different treatments
of theoretical uncertainties will give different conclusions on
this point.

Let us examine one of these quantities in a little more detail to
see how the theoretical uncertainties arise. In each case there is
a mix of weak interaction and short-distance strong-interaction
physics, which both are  perturbatively calculable and long range
strong-interaction physics which is not perturbatively calculable.
Tomorrow's lecture will introduce some of the methods that are
used to deal with (or avoid) possible long-range strong
interaction effects.  Here I simply want to show how such effects
can enter. Consider the question of the mass difference between
the two mass eigenstates for $B_d$.  The two one-loop diagrams
given in Fig. \ref{fig:1} are the dominant contribution to this
effect. Each loop-diagram can have either a $t$-, $c$-, or
$u$-quark for each of the two internal quark lines. Calculation of
the matrix element of these diagrams between a $B^0$ and a $\bar
B^0$ meson would give $M_{12} +i\Gamma_{12}/2$.

The diagrams can be written as a local four-quark operator
multiplied by a calculable coefficient which includes CKM factors.
I will write the quark-propagator and coupling dependent part of
this coefficient schematically as
\begin{equation}
Q =|V_{td}V_{tb}^*D_t + V_{tcd}V_{cb}^*D_c +V_{ud}V_{ub}^*D_u|^2
\end{equation}
where the $D_q$ factors are the quark propagators. This expression
is schematic because in writing it as a perfect square I ignored
the differences in the momenta of the two quark lines in the
diagram (which are typically small, ${\cal{O}}(m_b/m_W)$, compared
to the loop momentum itself).

Notice that if all the quarks had equal mass then $D_t=D_c=D_u$
and the unitarity condition Eq.~(\ref{unitarity}c) would say that
this factor $Q$ vanishes. Indeed we can use this condition to
rewrite the expression as
\begin{equation}
Q = |V_{td}V_{tb}^*(D_t-D_u) + V_{cd}V_{cb}^*(D_c-D_u)|^2.
\label{Q2}
\end{equation}
Because of the two $W$-propagators the loop integral is dominated
by momenta of order $M_W$, which is large compared to either the
$c$ or $u$ quark masses.  Thus the two quark propagators in the
second term of Eq.~(\ref{Q2}) above essentially cancel
one-another, so the term is suppressed by a factor of order
${(M_c^2 -M-u^2)}/{m_W^2}$.  Thus the mass difference is
effectively proportional to the square of the coefficient of the
remaining term, which $|V_{td}|^2$ (since $V_{tb}$ is 1 up to
order $|\lambda|^4$).  (Note that this argument also shows why the
mixing matrix is small in the $D$-meson case. There the three
propagators are the down-type quarks, all three of which have
masses that are small compared to $M_W$, so the Unitarity
cancellations suppress the entire effect. Furthermore the
contribution of the most-massive quark in this case, the
$b$-quark, is Cabibbo-suppressed, further reducing the effect. )

 To find the value of this $V_{td}$ by measuring the $B$ meson
mass differences we need to know the matrix element of the four
quark operator between the $B^0$ and $\bar B^0$ meson states. This
is where the long-distance hadronic physics sneaks into the
problem, this matrix element depends on the form of the $B$
wavefunction, including all effects of soft gluons.  The best
available method to determine it is to use lattice QCD calculation
\cite{latticevtd}.

A measurement of the mass difference of the two $B_d$ mass
eigenstates thus gives a measurement of $V_{td}$ with a
theoretical uncertainty that is dominated by the theoretical
uncertainty in the lattice determination of the relevant
four-quark matrix element. The result is usually written as some
``known'' factors times $B_B f_B^2$. (The ``known'' factors
include quark masses, which are actually not so well-known and
must be carefully defined.)   Here the factor $f_b^2$ is the
vacuum to one meson matrix element of the axial current which
arises in the naive approximation to the matrix element obtained
by splitting the four-quark operator into two-quark terms and
inserting the vacuum state between them.  This is known as the
vacuum-insertion approximation.  The quantity  $B_B$ is simply the
correction factor between that approximate answer and the true
answer. It can be estimated in various model calculations.  The
lattice calculation does not need to make this subdivision, it
directly calculates the full matrix element. However the result is
often quoted in terms of the $B_B$ and $f_B$ parameters. Lattice
methods can also directly calculate the latter. Eventually $f_b$
will be measured and that will provide a separate test of the
lattice calculation.

Once there is a good measurement of the $B_s$ mass difference the
ratio $\Delta m_b /\Delta m_s$ will provide a better determination
of $V_{td}$ via the ratio $V_{td}/V_{ts}$.  This mass ratio is
relatively free of theoretical uncertainties, as most of these
cancel in the ratio of matrix elements. The matrix elements for
the $B_d$ and the $B_s$ mesons are
 similar.  Only a small correction
due to the difference of the $s$ and $d$ quark masses remains. The
uncertainty in this correction  gives a relatively small
theoretical uncertainty in $V_{td}$.  At present  only a lower
limit for the $B_s$ mass difference is known; even this gives an
important constraint (upper limit) on the range of $V_{td}$.

\boldmath
\subsection{Time Evolution of the $B$ States and Time-Dependent
\hfill\break  Measurements} \unboldmath

Now I turn to the topic of  decays of neutral $B$ mesons.  What
can we measure and what does it tell us? To discuss this we need
to understand the time evolution of state which at time $t=0$ is
known to be a pure $B^0$ meson.  This means that at t=0 we have
\begin{equation}
B(t=0) = (B_H + B_L)/2p \ .
\end{equation}
Since the two mass states evolve with different time-dependent
exponential prefactors we find
\begin{equation}
B(t) =g_+(t)B^0 +(q/p)g_-(t)\bar B^0
\end{equation}
where the functions $g_\pm$ are just the sums and differences of
the exponential mass and lifetime factors
\begin{eqnarray}
g_\pm &=& [e^{(-iM_Ht-\Gamma_Ht/2)} \pm e^{(-im_Lt
-\Gamma_Lt/2)}]/2\nonumber \\
&=& e^{-iMt-\Gamma t/2}[e^{(-i\Delta M -\Delta \Gamma/2)/2} \pm
e^{(i\Delta M +\Delta \Gamma/2)/2}]/2 \ . \label{gplusminus}
\end{eqnarray}
Here we introduce the notation $M$ and $\Gamma $ for the average
mass and width and $\Delta M$ and $\Delta \Gamma$ for the
differences between the two sets of eigenvalues.  In the case of
$B_d$ the width difference is small compared to the mass
difference (and to the width itself) so to a good approximation we
can neglect $\Delta \Gamma$.  Then the expressions for the $g_\pm$
simplify in an obvious way.  For $B_s$ it is likely that the width
difference is comparable to the mass difference and the full
expressions must be used.

The time-dependent state that is a pure $\bar B^0$ at $t=0$ can
likewise be written in terms of these same functions
\begin{equation}
\bar B(t) = (p/q) g_-(t)B^0 +g_+(t)\bar B^0.
\end{equation}
It is now straightforward to derive the time-dependent rate to
reach a particular $CP$ eigenstate final state $f$ with $CP$
quantum number $\eta_f$. It is given by
\begin{equation}
|A(B(t)\rightarrow f|^2 = |A(B^0\rightarrow f)|^2[|g_+(t)|^2 +
|\lambda_f g_-(t)|^2 + 2Re[g^*_+(t)g_-(t)\lambda_f]]
\end{equation}
where the quantity
\begin{equation}
\lambda_f = (q/p) {A(\bar B \rightarrow f)\over A(B\rightarrow f)}
=\eta_f (q/p) {A(\bar B \rightarrow \bar f)\over A(B\rightarrow
f)}.
 \end{equation}
In the second equality here we have used the fact that f is a $CP$
eigenstate,
 $ CP f =\bar f = \eta_f  f$ where $\eta_f = \pm 1$,  to write the ratio of
amplitudes in a form that shows explicitly that one amplitude is
simply the $CP$ conjugate of the other.

The $CP$-violating asymmetry between the rates is defined to be
\begin{equation}
a(t) ={|A(\bar B(t)\rightarrow \bar f)|^2 -|A( B(t)\rightarrow
f)|^2\over |A(\bar B(t)\rightarrow \bar f)|^2+|A( B(t)\rightarrow
f)|^2}\ . \label{asymdef}
\end{equation}
(Note once again you must beware of conventions, some of the
literature defines the asymmetry with the opposite sign.)

If $\Delta \Gamma/\Gamma$ can be neglected, which is a very good
approximation for $B_d$ decays, then $|q/p|=1$ and the asymmetry
takes the form
\begin{equation}
a(t) = -[(1-|\lambda_f|^2)\cos(\Delta M t)+ 2Im\lambda_f
\sin(\Delta Mt)]/(1+|\Lambda_f|^2) \ . \label{asym}
\end{equation}
As promised previously, this relationship shows that the
$CP$-violating effects measure properties of $\lambda_f$, in
particular its magnitude and imaginary part.  (In the more general
case the expressions are somewhat more complicated and depend also
on the width difference.)  In particular, if only the third type
of $CP$ violation is present, namely if in addition to $|q/p|=1$
we have $|\bar A/A|=1 $ so that $|\lambda_f|=1$, then this
expression simplifies to
\begin{equation}
a(t) = -Im \lambda_f \sin(\Delta Mt)] \ .
\end{equation}
The argument of $\lambda$ depends simply on weak phases, so that
\begin{equation}
Im\lambda_f = \eta_f \sin(2\phi_{\rm mixing} -2\phi_{\rm decay})\
. \label{imlambda}
\end{equation}
Here $2\phi_{\rm mixing}$ is the phase of $q/p$ and $2\phi_{\rm
decay}$ is the phase of $A(\bar B \rightarrow \bar f)/A(B
\rightarrow f)$ while $\eta_f$ is the $CP$ quantum number of the
state $f$. These phases are each given by some combination of
$CKM$ matrix-element phases. While each of them separately can be
changed by changes in phase convention (rephasing of quark fields)
the difference is convention independent, as must be so for any
physically measurable quantity. Thus the asymmetry directly
measures the phase differences between particular CKM matrix
elements with no uncertainties introduced by our inability to
calculate strong interaction physics effects such as the magnitude
or strong phase of an amplitude. These strong interaction effects
all cancel exactly when  $|\lambda_f|$ is 1.

\boldmath
\subsection{CP Eigenstate Channels for $b\rightarrow c \bar c s$}
\unboldmath

There are many possible channels to investigate.  The interest
lies not just in one measurement but in whether the pattern of
$CP$-violating asymmetries fits the predictions of the Standard
Model. What channels should we study?  We need a final state of
definite $CP$.  In general for a multibody final state even when
the particle content is $CP$-self conjugate there will be an
admixture of $CP$-even and $CP$-odd contributions
 because of different possible orbital angular
momenta among the particles.  The simplest way to get a definite
$CP$ final state is to require that the $B$ decay to a two-body or
quasi-two body final state with only one allowed orbital angular
momentum.  (Quasi-two-body here simply means a two-body state with
one or two unstable particles, such as a $\rho \pi$ or $\rho
\rho$. The actual observed final state is then three or four
pions.) Given that the $B$ has spin zero, the final state has a
unique orbital angular momentum between the pair of particles if
(and only if) at least one of the two particles has spin zero.
For quasi-two body states where both particles have non-zero spin
but at least one of them is unstable one can possibly separate out
the $CP$-even and $CP$-odd final state contributions using an
angular analysis of the distribution of secondary decay products
\cite{anganal}. The price is that, in general, a larger data
sample is needed to achieve the same accuracy on the $CP$
asymmetry measurement.

Note that the Feynman diagram structure is the same for all
channels with the same quark content. Results from multiple
channels can sometimes be combined to improve statistical
accuracy.  For example for the quark decay $b\rightarrow c\bar c
s$ the $B^0$ decay channels $J/\psi K_S, \psi^\prime K_S,\eta_c
K_S$ $J/\psi K_L, \psi^\prime K_L,\xi_c K_L$ ({\em etc.}) all
depend on the same set of quark diagrams.  For the $b\rightarrow
u\bar u d$ (and $d\bar d d$)quark content there are likewise many
channels: $\pi\pi,\rho\pi, \rho\rho$, {\em etc.}  (The last of
these needs angular analysis.)

Let us then examine what the predicted $CP$ asymmetry is in each
of these two cases.  We begin with the modes such as $B\rightarrow
J/\psi K_s$. These have been called the golden modes for analyzing
$CP$ violation in $B$ decay.  For once we have a situation where
the mode for which the theoretical analysis is straightforward is
also one with good experimental accessibility.
 One still needs a large sample of $B$ decays
because the branching fraction to these channels is not large. (In
$B$ decays there are so many open channels that branching
fractions are small and smaller:  the ``large'' modes occur at the
few percent level; $J/\psi K_S$ and similar modes are about a
tenth of a percent; a ``rare'' mode in this game has a branching
fraction a few times $10^{-5}$.)

\begin{figure}[htb]
\centerline{\hbox{ \psfig{figure=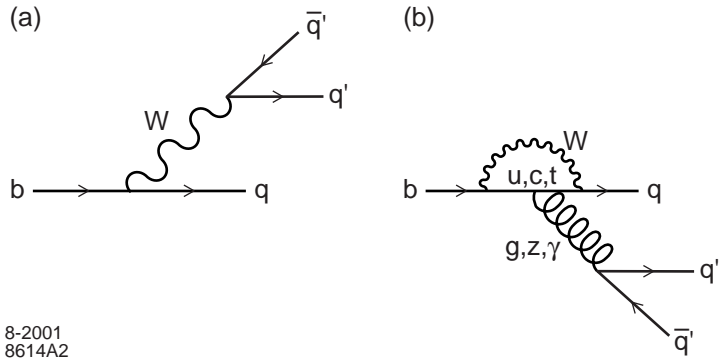}}} \caption[*]{
\footnotesize The (a) tree and (b) penguin weak decay processes at
the quark level.} \label{fig:2.1}
\end{figure}

First we need a little terminology. We use the term spectator
quark for the quark other than the $b$-type quark (or antiquark)
that is present in the initial $B$ meson, since it is generally
not involved in the $b$-decay diagram. There are two topologies of
weak decay Feynman diagram that can contribute to $B$ decays to
leading order in the weak interactions.  These are called ``tree''
and ``penguin'' diagrams and are shown in Fig. \ref{fig:2.1}.  A
tree diagram is one where the $W$-boson creates or connects to a
 different quark line from the line that starts out as the $b$-quark.
 I thus also include any annihilation diagram or any diagram where
the $W$-boson connects to the spectator quark as part of what I
call the tree amplitude. Whenever such a diagram is allowed it
will enter with the same CKM factors as the other tree diagram
processes. A penguin diagram is a loop-diagram where the $W$
reconnects to the quark line from which it was emitted. Then a
hard gluon is emitted from the quark line in the loop, and either
makes a pair or is absorbed by the spectator quark.

When higher order strong interaction rescattering effects are
included the distinction between tree and penguin
 diagrams becomes blurred. However, it is useful (and standard) to
start out by describing processes in this language as it allows us
to identify all the relevant CKM factors, and the operators which
they multiply. As we will shortly see, that is the essence of the
story. Eventually we will group terms not by the diagrams, but by
the CKM factors. That grouping is not blurred by any subsequent
strong interactions.  The language tree and penguin persists, but
the ``tree contribution", in my terminology will be taken to
include not only the tree diagrams (including those that involve
the spectator in the weak vertex), but also that part of the
contribution from the penguin diagrams that has the same CKM
factor as the tree diagrams. Obviously, if one wants to try to
calculate the size of the contribution to the amplitude one  must
keep track of each diagram separately, but if we are only
concerned with whether there is more than one CKM structure in the
significant contributions we can lump together all the terms with
a given CKM factor.

The cleanest cases theoretically  are those where we can make a
prediction without knowing anything about the sizes of the
amplitudes because we are looking at a ratio of rates where these
cancel to a good approximation.  The $CP$-violating asymmetry in
channels arising from quark transition $b\rightarrow c\bar c s$ in
a $B_d$ meson is just this type.  The tree diagram has a CKM
factor $V_{cb}^*V_{cs}$.  Any time that penguin diagrams
contribute to an amplitude there are three terms, corresponding to
the three different up-type quarks that inside the loop.  Thus we
can write the $b$ to $s$ penguin amplitude $P$ in the form
\begin{eqnarray}
P&=&V_{tb}^*V_{ts} f(m_t)+V_{cb}^*V_{cs} f(m_c)+V_{ub}^*V_{us}
f(m_u)\nonumber \\ &=&V_{cb}^*V_{cs}[
f(m_c)-f(m_t)]+V_{ub}^*V_{us} [f(m_u)-f(m_t)] \label{cspenguin}
\end{eqnarray}
where the $f(m_q)$ is some function of the quark mass.  In the
second expression I have once again used the Unitarity
relationship Eq.~(\ref{unitarity}c) to rewrite the three terms in
$P$ in terms of two independent CKM factors.  Notice that the
first of these is the same as that for the tree term, so for this
discussion we call that contribution part of the ``tree
amplitude''. The remaining term is CKM suppressed by an additional
factor of $\lambda^2$.  The two differences of
quark-mass-dependent factors are expected to be comparable in
magnitude. Furthermore, ignoring CKM factors, the penguin graph
contribution is expected to be suppressed by about 0.3 compared to
the tree graph, because it is a loop graph and has an additional
hard gluon. This means the suppressed second term in
Eq.~(\ref{cspenguin}) is negligible (a few percent) compared to
the ``tree amplitude'' which here is the sum of the tree term and
the dominant penguin term.

Thus
 we have an amplitude that effectively has only a single
CKM coefficient and hence one overall weak phase.  This then
ensures $|\bar A/A|=1$, which means there is no  decay-type
(direct) $CP$ violation. (You will recall we needed two terms with
different weak phases to get such an effect. )   Remember too that
for $B_d$ we expect $|q/p|=1$ to a good approximation.  Thus we
have a case where $|\lambda_f|=1$ and the measured asymmetry
arises purely from the interference of decay before and after
mixing. We find
\begin{equation}
a_{J/\psi K_S} = -Im (\lambda_{J/\psi K_S}) \sin (\Delta M t )=
\sin(2\beta)\sin (\Delta M t ) \ .
\end{equation}
Here the quantity $\beta$ is the lower left-hand angle in the
standard $B$ physics Unitarity triangle (also sometimes called
$\phi_1$). (The minus sign disappears because $\eta_f=-1$ for
$f=J/\psi K_S$.)  Thus this asymmetry directly measure the phase
of a rephasing-invariant combination of CKM elements.

Furthermore all the channels in the $c\bar c s$ list above measure
the same asymmetry, up to an overall sign, the $\eta_f$ factor of
the channel in question.  For example $K_S$ and $K_L$ are states
of opposite $CP$, as are the $\psi$ and $\eta_c$.  Care must be
taken to include the correct $\eta_f$ factor for each state in
combining the results.  One can also include a state such as
$J/\psi K^*$ provided the $K^*$ decays to a flavor-blind
combination such as $K_S\pi^0$, and angular analysis is used to
separate $CP$-even and $CP$-odd contributions.

One can apply this same diagrammatic analysis to the decays
$b\rightarrow c\bar cs$ in a $B_s$ meson.  This gives a prediction
for channels such as $J/\psi\phi$ that the $CP$ asymmetry is zero
in the Standard Model, as the $B_s$ mixing term is dominated by
CKM factors with the same weak phase as this decay.  Thus, in the
Standard Model, only the CKM suppressed penguin terms which we
neglected above can give $CP$ violating asymmetries here, so at
most a
 few percent asymmetry is expected.  Such
predictions of small or vanishing asymmetries give another way to
examine the patterns of the Standard Model.  Any theory of new
physics effects which give additional mixing contributions could
destroy the cancellation of mixing phase and decay phase which
makes this asymmetry small in the Standard Model. However to
interpret such a result one indeed needs some calculation of decay
amplitudes, in order to quantify more precisely how big the ``few
percent'' Standard Model asymmetry could be.

The trick of rewriting the sum of three penguin terms as two terms
using the Unitarity relationships is a generally useful tool.  In
any channel one then has at most two CKM factors to consider. The
next step is to get a rough estimate of the relative size of the
two terms. This becomes important when $|\bar A/A|\neq 1$.

\begin{figure}
\centerline{\hbox{ \psfig{figure=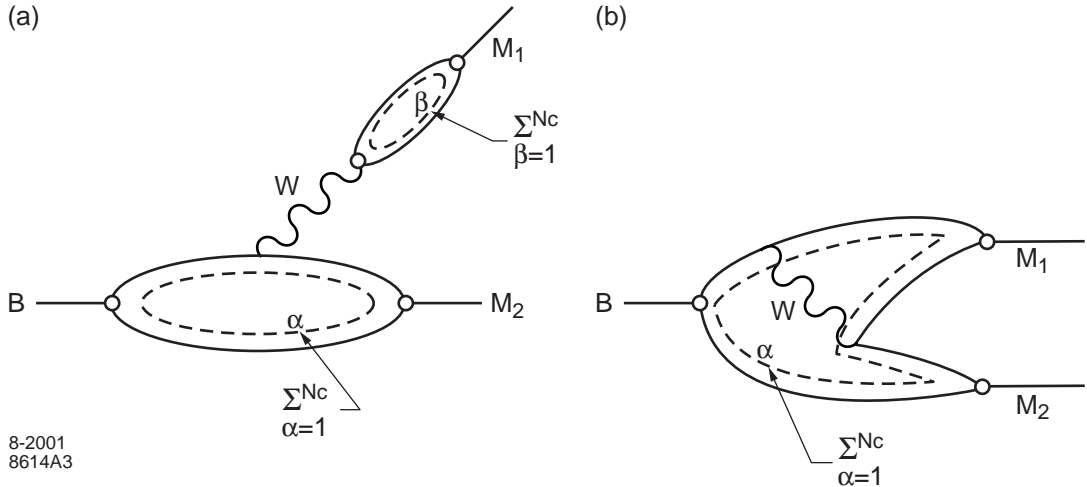}}}
\caption[*]{\footnotesize Possible two-meson tree-diagram decay
processes showing color-flow loops as dotted lines.  These are
called (a) color-allowed tree contribution, and (b) color
suppressed tree contribution.} \label{fig:2.2}
\end{figure}

\boldmath
\subsection{Some further $B$ Physics Jargon}
\unboldmath

The $B$ physics jargon  distinguishes contributions by three
attributes, because these three things give a first estimate of
how big the contribution is. The first size factor is whether the
diagram is tree or penguin. The penguin is suppressed relative to
the tree because it is a loop diagram and because it involves a
factor of $\alpha_{\rm strong}$ at a scale of order $m_b$ due to
the hard gluon, together this makes for a suppression factor of
order about 0.3, all else being equal. The next size factor is the
powers of the Wolfenstein parameter $\lambda$ in the associated
CKM factors.  All $B$-decay amplitudes have at least two powers of
$\lambda$.  Amplitudes with higher powers are called
CKM-suppressed.  The third size factor is the color flow pattern
that forms the particular final state of interest.  Diagrams where
a quark-antiquark pair produced by a W finish up in the same meson
are called color-allowed, because this pair is produced in the
requisite color-singlet combination.  In terms of color-flow
diagrams there are two independent color-flow loops as shown in
Fig. \ref{fig:2.2}(a).  When the quark and antiquark produced by
the $W$ end up in different final mesons the diagram is called
color-suppressed (Fig. \ref{fig:2.2}(b)). There is then only a
single color-flow loop so that diagram is expected to be of the
order of $1/N_c$ smaller than the corresponding color-allowed
diagram.

\begin{figure}[htb]
\centerline{\hbox{ \psfig{figure=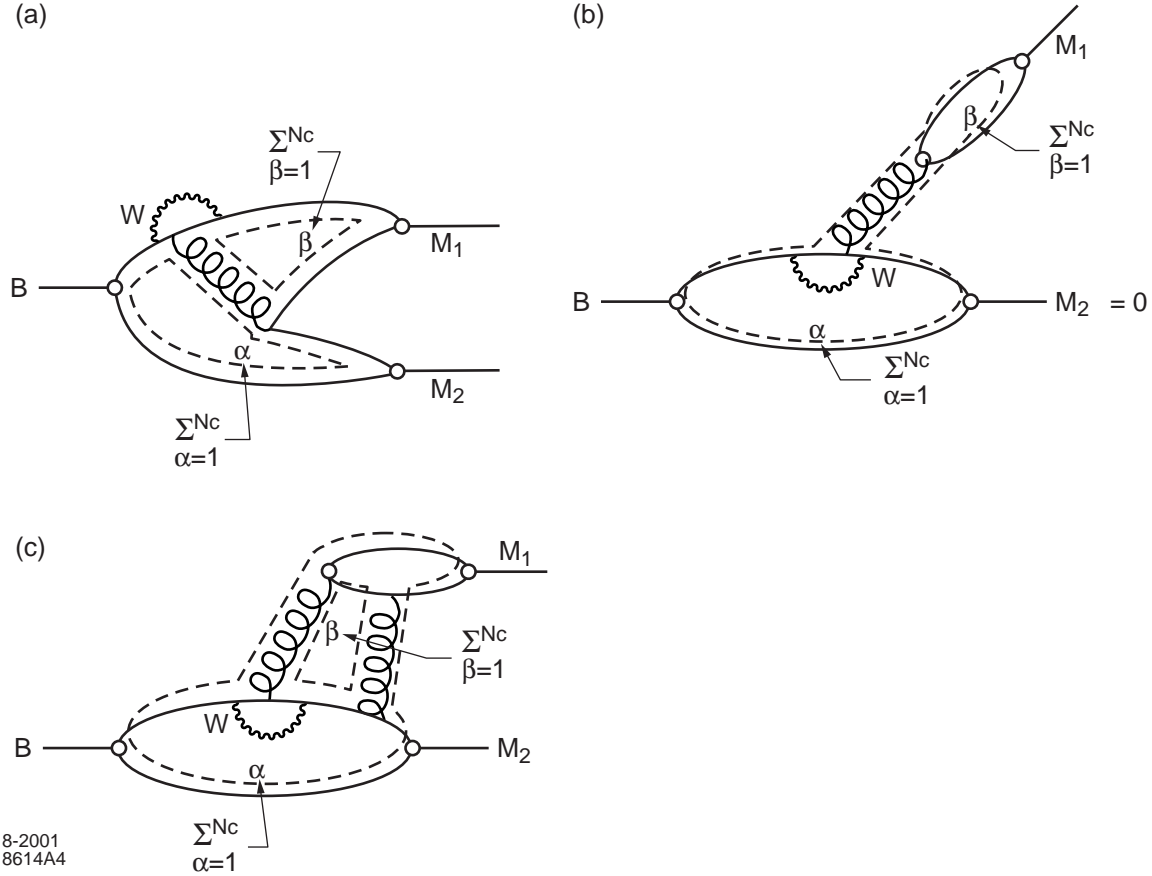}}}
\caption[*]{\footnotesize Possible penguin-type two-meson decay
processes showing color-flow loops as dotted lines.  These are
called (a) color allowed penguin, (b) naive color suppressed
penguin process, vanishes exactly, and (c) allowed diagram with
additional gluon for so-called color-suppressed penguin process.
(It has two color flow loops as does the ``color-allowed'', but an
additional $\alpha_{qcd}$ factor.)} \label{fig:2.3}
\end{figure}

 For penguin diagrams color suppression, if
it works at all, works the other way around.  Diagrams where the
quark and antiquark from the gluon end up in two different mesons,
Fig. \ref{fig:2.3}(a), are color allowed, and indeed can be seen
to have two-color-flow loops just as do the tree color-allowed
contributions.  Diagrams where the flavor-structure says the quark
and antiquark produced by the hard gluon must be in the same meson
are called color suppressed.  In Fig. \ref{fig:2.3}(b) there is
only one color loop.  However in this diagram the gluon makes a
color singlet object. But a gluon is a color-octet state. Taken
literally, the diagram vanishes. A second gluon must be exchanged
here.  If we were to count the extra gluon as a hard gluon, there
would be an additional suppression factor of $\alpha_{\rm
strong}$, but no $1/N_C$, because we would again see two color
loops, Fig. \ref{fig:2.3}(c). However the second gluon is not
necessarily hard, so the relevant scale for the $\alpha_{\rm
strong}$ is not large.  In some estimates these contributions are
treated as $1/N_C$ suppressed terms, but there is no good argument
that justifies this counting. As you can see from these arguments,
the naive color-counting is not a very reliable measure of the
relative strengths of the two types of penguin contributions.
QCD-improved operator-product expansion calculations at leading
order in $\Lambda/m_b$ \cite{qcda,qcdb,qcdc} can be made. These
treat the color factors correctly.  We will return to this
approach at later, in Lecture 3. However there is a large
literature of estimates that use the language of color-allowed and
color-suppressed contributions, so it is important to know how
these terms arose and how they are used.

All these size-counting factors are generally used to give first
estimates of the order of magnitude of the various contributions.
Clearly a more serious calculation can significantly change the
relative sizes. The kinematics of the different diagrams are
different. The matrix elements of the various operators are
different.
  Indeed there is an interplay between the wave
function of the mesons and the counting factors discussed above
which in the end determines the size of an amplitude.  Powers of
$\Lambda_{QCD}/m_b$ can arise from the wavefunction for particular
kinematic configurations relative to others.  Higher-order hard
QCD effects can be systematically included, but the soft
hadronization part of the calculation needs some additional input,
either from a model or from some other measurement.

\subsection{Another Sample Channel}

Now let us look at one more set of channels to see what happens
when this size counting  says two CKM factors can occur with
comparable coefficients. The case I choose to examine is the decay
$B_d \rightarrow \pi^+\pi^-$.  At the quark level this process is
governed by decays $b \rightarrow u\bar u d$. You can readily find
from the diagrams of Fig.~\ref{fig:2.1} that there are both tree
and penguin contributions for this quark content. The tree
diagrams have a CKM factor $V_{ub}^*V_{ud}$.  For the penguin
contributions we can again use unitarity to rewrite the three
different intermediate quark contributions
 as a sum of two terms.  In
this case all three CKM coefficients are of the same magnitude. I
choose to eliminate $V_{cb}^*V_{cd}$ because then the second
penguin term (the one that does not have the same weak phase as
the tree term) has the same weak phase as the mixing term in the
Standard Model. Then only one difference of CKM phases will enter
my eventual formulae for the asymmetry.  However we cannot ignore
the second penguin term.
 The only thing that makes it small compared to the ``tree amplitude''
(which includes the first penguin term as well as the contribution
from the tree diagram) is the fact it is a penguin loop. That is
not sufficient to completely discard it.

So here we have a situation where there can be $|\bar A/A|\neq 1$
effects.  We must use Eq.~(\ref{asym}) to interpret the  the
measured asymmetry. One would like to extract from the measurement
the CKM phase difference between mixing and tree decay
contribution (which in this case is $\alpha \equiv \pi -\beta -
\gamma$).  One can measure two quantities, $|\lambda_f|$ from the
coefficient of cos($\Delta M t$), and Im$\lambda_f$ from the
coefficient of sin($\Delta M t$).

However three unknown quantities enter in the expressions for
$\lambda_f$ in such a case.  These are the relative weak phase of
mixing and the tree decay amplitude $\alpha$, and both the
absolute value ratio, r, and the relative strong phase, $\delta$
of the penguin and tree terms. We can write
\begin{equation}
\lambda_f = e^{-2i\alpha}{1 +re^{i(\delta +\alpha)} \over
1+r^{i(\delta -\alpha)}} \ .
\end{equation}
Here the phase $\alpha = \pi - \gamma -\beta$ is the angle at the
top vertex of the standard $B$-physics unitarity triangle; it is
the difference between the weak phases of the mixing and that of
the tree contribution to the decay. Obviously, knowledge of both
the real and imaginary parts of $\lambda_f$ is not enough to fix
all three quantities. So we cannot extract a value of $\alpha$
from this asymmetry measurement alone. (Note, however that for
very small r the expression simplifies so that the measurement of
Im$\lambda$ determines sin$2\alpha$.) We must use further theory
or measurement inputs (or both) to determine $\alpha$ if r is not
small.  (A note of warning here, one often sees the statement that
one tests the Standard Model by testing the relationship $\alpha =
\pi -\beta -\gamma$ between the angles in the triangle.  The
relationship is a definition. The tests of the Standard Model are
tests of whether one finds the same result for the two independent
angles, usually chosen to be $\beta$ and $\gamma$, using a variety
of independent ways to measure them.)

Note also that the ratio, $re^{i\delta}$, of the tree to the
penguin amplitudes will be different for the different channels
with the same quark content.  The kinematics of the tree and
penguin diagrams are different, and so are the wave functions for
forming a $\pi$ or a $\rho$, for example.  Thus, unlike the $c\bar
c s$ decays, we cannot simply combine channels to improve
statistical accuracy. Instead we must devise methods to remove the
dependence on the additional parameters; these methods are
different for each set of final state particles.

For the $\pi \pi$ case there are two ways to proceed. One is to
rely on isospin symmetry and isospin-related channels to give the
needed additional information. The second is to  develop methods
to calculate these various amplitudes more reliably.  This may
also involve using relationships to other channels where the tree
and penguin amplitudes enter with different relative strengths
because of different CKM structure.  For example by using
measurements on $K\pi$ channels as well with those from $\pi \pi$
channels one can gain some information on the size of the penguin
amplitude which dominates the decay in the former case. One can
then use SU(3) symmetry to relate that to the size of the penguin
in the $\pi\pi$ case.  Eventually such methods can much reduce the
theoretical uncertainty in the extraction of the CKM parameter
$\gamma$, or equivalently $\alpha = \pi -\beta -\gamma$. Tomorrow
I will discuss both of these approaches in a little more detail.

The set of all possible $B$ decays can be summarized by reviewing
all possible $b$-quark decays and the channels to which they can
contribute. A little care must be applied to this logic, as strong
rescattering can turn one quark-antiquark combination into
another, one must include this possibility in a full treatment.
For example in any channel involving a $\pi^0$ or $\rho^0$ meson
the penguin diagrams for $b\rightarrow d\bar d d$ must be added to
the diagrams for $b\rightarrow u\bar u d$. I refer you to the
table in the Particle Data Book review on this topic
\cite{pdgreview} that summarizes the quark decays and gives the
CKM factors that enter for each (after using the Unitarity trick
to get two terms only.) Any time you start thinking about a
specific process you will find you want this information. You can
rederive it readily by drawing the allowed quark diagrams and
investigating their CKM factors.

\boldmath
\section{Lecture 3. Theorist's Tools for $B$-physics}
\unboldmath

Today's lecture will briefly introduce a number of theoretical
tools for calculating $B$ decay processes. There are only a few
examples of measurements for which we do not need to know the
relative magnitude of various contributions to the decay
amplitudes in order to relate the measurement to some parameters
in the theory. We would like to go further and interpret the
multitude of other measurements that are possible because of the
many different $B$-decay channels.  To do this we must devise
methods to calculate or relate amplitudes.  The available
calculational methods all involve some mix of systematic expansion
in powers of one or more small parameters, lattice calculation of
matrix elements of operators, relationships based on symmetries of
the strong interactions such as isospin and SU(3) flavor symmetry,
and some input for transition matrix elements and or quark
distribution functions.  These last can be calculated reliably
only in certain limits and in general require models and
approximations. Alternately one can measure some of these
quantities in one set of processes and use the measured values as
input in the interpretation of other measurements.

This lecture will give a general picture of the toolkit of
approaches, what each tool is, and how it can be used.  There will
not be time here to teach the details of any of the methods.  This
lecture summarizes a large body of theoretical work.  I will not
attempt to reference all the relevant papers, but will include
references to some current papers as examples of the type of work
now underway.  I apologize in advance to the many whose papers I
do not mention.

There are two small parameters in this game, namely
$\Lambda_{QCD}/m_b$ and $\alpha_{\rm strong}(m_b)$.  Here $m_b$ is
the mass of the $b$-quark and $\Lambda_{QCD}$ is the scale that
defines the running of the strong interaction coupling.  The
detailed definition of each of these quantities is fraught with
technical problems, but there is a clear physical meaning for the
rough size of these parameters. $\Lambda_{QCD}$ is related to the
inverse size of a typical hadron while the $b$-quark mass can be
characterized as roughly the same scale as the mass of a $B$ meson
(up to corrections of order $\Lambda_{QCD}/m_b$). The strong
coupling $\alpha_s(m_b)$ scales as a logarithm of
$\Lambda_{QCD}/m_b$; we treat it as a separate small parameter
because we can count powers of this parameter separately from the
powers of $\Lambda_{QCD}/m_b$; they arise in different ways.

The fact that $\Lambda_{QCD}/m_b$ is indeed quite small leads to a
simple intuitive picture of a $B$ meson at rest. It is an
essentially static $b$ quark with the light quark forming a cloud
around it. The light-quark distribution is sometimes called the
brown muck, because we cannot reliably calculate the details of
it.  However we do know that  certain properties  are rigorously
true in the limit $m_b\rightarrow \infty$.  For example in that
limit the wavefunction does not depend on the spin orientation of
the $b$-quark and hence is the same for a spin 0 $B$ meson and a
spin 1 $B^*$.  A second way in which the large mass of the
$b$-quark simplifies the problem is that any gluon that carries
off a significant fraction of the $b$-quark mass is a hard gluon
that can be treated perturbatively; it introduces the small
parameter $\alpha_{\rm strong}(m_b)$.

In addition to these expansions there is another part of the
picture that is true because $m_b/M_W$ is small. This means that
weak decays of the $b$-quark are essentially local four-quark
effects.  Thus the $B$ meson decay can, to a reasonable
approximation, be thought of as proceeding in two stages:  a
$b$-quark decays and then the remnants hadronize to give the final
state under study.  It is this second stage, the hadronization,
that introduces all the uncertainties into the calculations.  We
have good methods for applying QCD to things like jet-formation
for well-separated high momentum quarks, but a $B$ decay does not
give us large enough quark momenta to use this formalism reliably.
Further, we want to know amplitudes for specific few-body
(quasi-two-body) final states (states of definite $CP$).  Most
likely these arise when the four quarks that are present after the
$b$ decay are not well-separated (so even if the $B$ mass were
much larger a jet calculation would not provide the answer). We
cannot calculate these amplitudes completely from first
principles.  So my purpose in this lecture is to review the tools
that we do have and how they can be used to minimize the
theoretical uncertainty on the extraction of the desired
quantities, such as CKM parameters, from experiment.

\subsection{Operator Product Expansion}

The operator product expansion is a way to formalize the
separation of hard or short-distance physics from soft or
long-distance physics.  It begins by rewriting the Feynman
diagrams into the form of local operators, defined at a given
scale, with calculable, scale-dependent coefficients.

First we look at all the tree and penguin Feynman diagrams for the
weak decay of the $b$-quark. Each can be written as a sum of four
quark operators with definite coefficients at the scale $M_W$.
This is the leading order operator product expansion.  There are
actually two types of penguin diagrams, those I mentioned earlier
that involve a gluon, and a second set called electroweak penguins
that involve a photon or a $Z$ particle emitted from the loop.
These last give an additional set of four-quark operators. At
first glance one might guess that
 the electroweak penguin  contributions are very small, with
$\alpha_{QED}$ replacing the $\alpha_{\rm strong}$ of the gluon
case. However it turns out there is a part of the $Z$-penguin
contribution which is enhanced by a factor $M_t^2/M_W^2$ and so
there are cases where these terms can be important too.

Each class of diagrams corresponds to a distinct set of four quark
operators at leading order.  When hard QCD corrections are
included, one must introduce a new scale into the problem, which
is the hard-soft separation scale $\mu$ that defines which gluons
are absorbed into the new scale-dependent operator coefficients
and which are defined to be included in the scale-dependent matrix
elements of operators. In addition, these corrections can mix the
operators, and thereby blur the distinction between tree and
penguin contributions.  Thus the labels of each operator as being
tree or penguin type is a leading order distinction only.  However
they are usually listed in that way as it is a useful way to keep
track of which operator arises with which CKM coefficients. In
addition, if a hard gluon connects the weak decay vertex to the
spectator quark this can also introduce additional local operators
that involve six quark fields, again with calculable coefficients
that begin at order $\alpha_s(m_b)$.

One must choose the $\mu$-scale that separates hard and soft
physics. In principle no physics depends on this choice.  In
practice if one makes approximations for the matrix elements one
does not usually get the correct scale-dependence in their values.
So  results do to some extent depend on the choice of scale.  This
dependence is minimized by doing higher order QCD calculations,
but in general is not fully removed even with that laborious step.

Each four-quark operator takes the form
\begin{equation}
{\cal O}_n = \bar b \Gamma_{n1} q^i \bar q^j\Gamma_{n1}q^k
\end{equation}
where each $\Gamma_{ni}$ denote a specific combination of gamma
matrices and QCD color structure and the $q^i$ denote the relevant
quark flavor (and color) content.  The details of the color and
flavor flow in the diagram can be read off once these operators
are written.  I do not include here the detailed list nor any
discussion of the coefficients.  That is available many places
\cite{texts}; my point here is not to discuss this well-developed
technical subject, but rather to talk about the additional steps
between writing down an operator and its coefficient and
calculating an amplitude for any particular channel.

The matrix elements of the operators between the initial $B$ state
and the final set of mesons are where hadronic physics enters the
game.  Our methods for calculating that physics are limited.  We
can however use information that we do have about symmetries of
the strong interactions, for example, to tell us about the ratios
of matrix elements that occur in different decays.

\subsection{The Factorization Approximation}

The simplest approach to the problem, for example for calculation
of a color-allowed tree diagram, is to approximate the matrix
element in a two-hadron decay as the product of the transition
matrix element of a two-quark weak current between the $B$ meson
and one final state meson (that can be measured in a semileptonic
decay), times the matrix element for the $W$ to create the second
meson, which is also measured elsewhere.  This approach is called
factorization, (or sometimes ``naive factorization'') because it
factorizes the four-quark hadronic operator matrix element into a
product of two two-quark matrix elements. This idea can be
generalized to divide any four-quark operator into two two-quark
operators, which can  either be extracted from experiment or
estimated using models for the quark distribution functions of the
mesons. The approximation neglects any effect of interactions
between the two mesons in the final state, effects known as final
state interactions.

Now we know that two mesons (for a concrete example think of two
pions) colliding at the energy corresponding to a $B$-mass
certainly do interact. So at first glance you may think this
approximation has no reason to be accurate.  It is certainly not
rigorously true, except in a few special cases.  However it is
motivated by a reasonable physical picture, usually attributed to
Bjorken \cite{bjfactorization} (although in this reference he says
the argument is common knowledge).

The idea is that the weak decay is a very local process which
converts one quark to three.  Only for the kinematic configuration
where two of these quarks (or rather one quark and one antiquark)
go off essentially together, with the third one recoiling in the
opposite direction, is there any significant probability that the
system will hadronize as a two-body final state. (All other
configurations are assumed to make multi-body final states, for
example by  fragmentation of the four final-state quarks.)
 In the special case that gives two-body states the quark
and anti-quark that travel together start out much closer together
in the transverse direction than the size of a typical hadron.
They get quite far from the region containing the other quark and
the ``brown muck" of the spectator quark before they evolve into
the hadronic-sized meson that is observed.  They must start out in
a color-singlet state to form such a meson.  In a local
color-singlet configuration (small compared to a meson) the strong
interactions must cancel.  So initially there are no strong
interactions because the pair is in a local color-singlet
configuration.  Later there is no strong interaction because the
two mesons are well-separated and strong interactions are a
short-range phenomenon.

The justification of the factorization approximation, as described
above, applies for a tree diagram with no direct involvement of
the other valence quark of the $B$ meson quark in the weak decay
vertex.   More generally one can try to factorize any four quark
operator (possibly after making a Fierz rearrangement to group the
relevant quark fields as flavor-flow dictates they must be grouped
to form the mesons of interest).  One then uses other
measurements, or possibly lattice calculations, to fix the two
two-quark matrix elements.  In the case of a color-suppressed
contribution, or one arising from a penguin diagram the
flavor-flow does not automatically match two color-singlet quark
pairings.  However, if a color-singlet meson is to be formed then
there must be a color-singlet piece of the amplitude, and for this
piece the factorization argument applies.

In some processes the flavor content of the final state allows a
contribution either from annihilation (in the case of a charged
$B$ meson) or from exchange of a $W$ between the two initial state
valence quarks (for neutral $B$'s).  Both processes are suppressed
in the heavy quark limit by the quark-mass dependence of the
wave-function at the origin (the $B$ to vacuum transition matrix
element of a local two-quark current).  These contributions are
typically neglected in rough estimates of two-hadron decay rates.

Despite all the caveats, the factorization approximation is
generally used to make first guess estimates of the sizes of
various partial rates. To determine the reliability of this
calculation one must look more carefully at what is being done
here.  I mentioned previously that the operator coefficients can
be calculated with hard QCD corrections taken into account.  This
introduces a scale dependence into their definition, the scale of
the separation between hard and soft corrections in QCD. This is
not a physical scale, but an arbitrarily chosen one, so the true
answer cannot depend on it.  Any scale-dependence in the
coefficients must be compensated by cancelling scale-dependence in
the matrix elements. But when we use measurement of a
semi-leptonic process to determine the matrix element there is no
reference to any hard-soft division scale; the measured quantity
is scale independent.  So we clearly have a problem, even in the
best cases, factorization cannot be quite correct.

The naive way to deal with this problem is to say it is reasonable
to pick a scale somewhere between $m_b/2$ and $2m_b$ since the
mass of the $b$-quark sets the typical momentum scale for the
quarks arising from its decay.  One then asks how the quantity in
question varies as one changes the scale within this range and
uses this variation to assign a central value and a theoretical
uncertainty to the result.  While this seems quite a plausible
approach there is no way to be sure it is right.  The problem is
alleviated somewhat, though not completely removed, when higher
order QCD calculations of the operator coefficients are used. It
can only be dealt with correctly when a consistent treatment of
higher order matrix elements is used, along with the higher order
coefficients. Any finite order calculation, however, will
typically have some residual scale-dependence problems.

The issue of determining the theoretical uncertainty, that is the
reasonable range of values of a theoretical estimate,  is one to
which we will return again and again in this lecture.  Our ability
to test the Standard Model by comparing its predictions with
experiment depends on our ability to determine how big the
uncertainties in our theoretical calculation are. A clean result
is one where we know that these uncertainties are very small, or
at least where we know very well how big they can be.  But more
often than not we find a part of the calculation is not so clean.
The methods of determining the possible range of the predictions
of the Standard Model are all too often subjective and
ill-defined.  Theorists continue to work to remove such
ambiguities, and to find those measurements, or sets of
measurements, for which they are minimal. This is an important
task.

\boldmath
\subsection{Heavy Quark Limit Relationships between $B$ and $D$
\hfill\break Mesons} \unboldmath

One powerful technique for dealing with $B$ decays is use the fact
that the $b$-quark mass is large compared to the QCD scale and to
calculate quantities in terms of a power series expansion in that
ratio.  If one also treats the charm quark as heavy compared to
the QCD scale then one has an even more powerful set of
relationships.  Then to leading order in $\Lambda_{QCD}/m_q$ the
distribution of the light quark in a heavy-light meson is
independent of the spin orientation or the mass of the heavy
quark.  This means it is the same for a $B$ or a $B^*$ or a $D$ or
a $D^*$ meson.  This is a very important statement because it
gives us at least one limit in which we know the transition matrix
element between a $B$ and a $D$ or $D^*$ meson.

Consider for example the semi-leptonic decay $B^0\rightarrow
D^*\ell\nu$. In the kinematic limit where the $D^*$ is at rest in
the $B$ rest frame the wave-function overlap is 1. There is a
small but calculable QCD correction to the unit wave-function
overlap. Then there are the corrections to the heavy-quark limit
relationships, which in this case turn out to be quadratic in
$\Lambda_{QCD}/m_q$.  This is reasonably small even for the charm
quark.  This means that we can, in principle, use a measurement of
this quantity to extract the CKM matrix element $V_{cb}$ with very
little theoretical uncertainty.   The only problem is that the
configuration where this relationship holds is, as I said, a
kinematic limit.  That means that the rate vanishes at that point!
One must measure the rate as a function of $q^2$, and use an
extrapolation to extract the quantity of interest.  The
extrapolation requires some knowledge about the behavior of the
form factor as one goes away from the perfect-overlap situation,
and that introduces some theoretical uncertainty into the answer
for $V_{cb}$.  However as more data is collected one can measure
the rate ever closer to the end point, thereby reducing the
sensitivity to the extrapolation.

There are some other technical issues that appear in this problem.
One interesting one that crops up here, and in other problems too,
is the choice of the definition of the quark mass $m_b$ (or
$m_c$).
 If you remember from muon decay, the semileptonic decay rate for a
fermion (here the $b$-quark) goes like the fifth power of the mass
of the decaying particle. Thus any uncertainty in the definition
of the quark mass translates into a huge uncertainty in the
predicted rate.  But it is even worse than this.  If you try to
define the quark mass as the mass at the pole of the quark
propagator this definition is scale dependent and even diverges as
the scale is reduced (known as the renormalon problem). Clearly
this is an unphysical effect, because you chose an unphysical
definition of the quark mass.  The problem is to find a definition
that avoids this problem and leads to a well-controlled result.
This can indeed be done.  The full discussion of how one does it
is beyond the scope of this lecture.  I merely warn you that you
can get into trouble by blithely assuming you know what someone
means when they write $m_b$.  This quantity cannot be directly
measured.  It is dependent on definition convention and on
renormalization scale.  As you compare results of different
calculations you must always be aware of the conventions and
definitions that have been used.  Otherwise you will not be able
to interpret and apply the results correctly.

\subsection{QCD-Improved Factorization}

The word picture explanation of factorization is to some extent
confirmed by explicit calculation of QCD corrections up to order
$\alpha_S$ and at leading order in $\Lambda/m_q$.  It is found
that the color-singlet nature of the meson leads to cancellation
of the  soft-gluon exchange between the two final-state mesons. In
general, particularly for processes dominated by penguin or
color-suppressed diagrams, there are found to be additional
contributions which cannot be described by the simple
factorization of a four-quark operator, but rather add to the
picture a local six-quark operator.  They arise because of a
hard-gluon exchange between the so-called spectator quark (now no
longer just a spectator) and another quark within the same meson.
The matrix elements of this operator can be approximated as the a
product of three valence-quark-distribution functions, one for
each meson (one initial and two final) times the hard coefficient
which begins in order $\alpha_s(m_b)$.  Uncertainties arise from
limitations on our knowledge of the quark distribution functions.

One has to be careful here when matching the calculated hard-quark
coefficient with measured transition matrix elements and form
factors. The scale-dependence matching must be done correctly. One
must also ensure that one is not double counting contributions of
hard quarks that are effectively inside one of the measured
quantities. But these are technical problems that can be dealt
with correctly.

This treatment is known as qcd-improved factorization \cite{qcda}.
Here the term factorization is used for the factorization of the
hard and soft physics. This form of factorization has been
demonstrated to work for the leading order in $\Lambda/m_b$ and
one order in $\alpha_s({m_b})$ corrections to the leading
diagrams. The actual $\Lambda/M_b$ power counting is dependent on
the assumptions about quark distribution functions; it assumes
they vanish as a power of x at their end-point. As the calculation
includes all gluon energy scales it is argued that all final state
interactions are included in the formalism.  The question remains
as to whether this argument applies to all orders. It has been
proven true to all orders in  $\alpha_s$ and leading order in
$\Lambda/m_q$ for the special case of a $D\pi$ final state with
flavor such that the spectator quark in the $B$ ends up in the $D$
and the charm quark is treated as a heavy quark in the
$\Lambda/m_q$ power counting \cite{dpifac}.

It turns out that the numerical results depend quite sensitively
on the details of input assumptions on the quark distribution
functions \cite{qcdb, qcdc}. A variant of the approach making
quite different, and indeed additional, assumptions about the
quark distribution function end-point behavior gets numerically
very different results \cite{qcdc}. The second approach is called
perturbative QCD by its proponents. It is claimed in this approach
that the entire result is perturbatively calculable. While these
claims are open to question \cite{sachrajda}, one can simply
 regard the results of this work as the output of a set of
 ansaetze for the distribution
functions. The results raise issues that have contributed
important points to the discussion. One is the question of exactly
how small some of the $(\Lambda/m_b)$-suppressed contributions are
in actuality. The annihilation-graph contribution, for example, is
found to be significant, even though formally suppressed.

The sensitivity of results to inputs is unfortunate. It means that
even these more sophisticated calculations leave us with some
significant theoretical uncertainties.  The best one can do to
quantifying these uncertainties is to see how much the results
change when one varies over some reasonable set of assumptions for
the various inputs such as quark distribution functions and
transition matrix elements.  But how do you decide what is a
reasonable range?  As the existing debates show, in many cases
this comes down to some subjective choices, not all rigorously
decidable! (Some choices are, however, quite clearly unreasonable
and should be excluded from discussion, for example a calculation
that sets the scale of transverse momenta in a hadron at
$k_\perp^2 =\Lambda m_b$, or a form-factor model that does not fit
a rigorous theoretical limit relationship.) As data and
calculations for multiple channels are obtained it is likely that
we will develop a better understanding of such issues, and a more
consistent view of what range of assumptions are reasonable will
emerge. Meanwhile it is very important that any calculation
reported should include an honest estimate of its uncertainties,
and a clear explanation of the assumptions made and the ranges of
input variables that were included in obtaining this estimate.

\subsection{Isospin}

Another useful tool for extracting clean results for strong decay
amplitudes is the symmetries of the strong interactions.  The best
of these, in that it most close to a true symmetry of the hadronic
decays, is Isospin symmetry.  I find I must explain this symmetry
from scratch for current students.  It is a piece of old fashioned
physics knowledge which is not always taught in modern courses.
Isospin is a symmetry under interchange of $u$ and $d$ quark
flavors.  It is called ``iso'', because atoms which differ by such
an interchange (originally by replacing a neutron by a proton or
vice versa) are called isomers because they have nearly equal
mass, and ``spin'' because the two quarks form an SU(2) doublet
and the mathematics of SU(2) is the familiar mathematics of spin
doublets.  Isospin has nothing to do with any angular momentum.
Notice also that I do not here mean the weak isospin (so called
because it is yet another SU(2)); the isospin doublet is truly $u$
with $d$, not with some admixture of $d$,$s$, and $b$.

Isopin is, quite obviously, broken by electromagnetic effects
since these distinguish quark charges, and it is also broken by
quark masses. Now the up and down quark mass are nowhere near the
same, the ratio $({m_u-m_d})/({m_u+m_d})$ is not a small number.
So why is Isospin ever a good symmetry?  The answer is that in
many cases, (including most but not all hadron decays) the
relevant scale with which to compare the quark mass difference is
not the quark mass sum but the hadron mass scale.  That scale is
set either by $\Lambda_{QCD}$ or by some heavy quark mass.  Then
the corrections to isopin-based predictions are small.  One must
be careful, however, to look out for the cases where the effect is
one that is ``chirally enhanced'' that is where the sum of up and
down masses does appear in the denominator.  (A similar issue may
also arise when making a heavy-quark expansion; terms that behave
like ${\Lambda_{QCD}^2}/{m_b(m_u+m_d)}$, though formally
suppressed in the large $m_b$ limit, are not always numerically
negligible.)

How does isospin help clarify $B$ decay processes?  Its chief
value is that it allows us to make an experimental separation of
some tree and QCD-penguin type contributions.  In some processes
these have different isospin structure, as well as having
different CKM structure.  Let us take the example of $B$ decaying
to two pions. First let us look at the final states, two pions in
a spin zero state.  A pion has isopin 1. Naively there are three
possible isospins for the two-pion states, 0, 1 and 2. However
Bose statistics says the overall state must be even under pion
interchange.  Since the spin zero spatial state is even, the
isopin state must be even too. This eliminates the $I=1$
possibility.  Now let us examine the quark decays.  The tree
$b\rightarrow u\bar u d$ contribution contains both $\Delta I =
1/2$ and $\Delta I = 3/2$ contributions.  These combine with the
spectator quark to contribute to the $I=0$ and $I=2$ final states
respectively.  But a gluon is an isosinglet particle---it has no
isospin.  Hence the $b \rightarrow d$ QCD penguin graph is purely
$\Delta I=1/2$ and contributes only to the $I=0$ final state. (In
quark language the gluon makes $u\bar u + d \bar d$. ) We can use
measurements of several isospin-related channels (Here $B^0
\rightarrow \pi^+\pi^-$, $B^0 \rightarrow \pi^0\pi^0$ and $B^+
\rightarrow \pi^+\pi^0$ and their CP conjugates) to isolate the
$I=2$ contribution \cite{gronaulondon}. Then we have found a pure
tree process, which thus depends on only one weak phase (up to
small corrections from electroweak penguin effects.) Thus the
isospin analysis gives us a way to separate out the dependence on
$\alpha$, the difference of the weak phase of the mixing and the
weak phase of the tree diagram, without having to calculate the
relative strength of the penguin and tree contributions.

The theoretical uncertainty that we found in the previous lecture
in trying to extract the CKM parameter $\alpha$ from the asymmetry
in $B \rightarrow \pi^+\pi^-$ decays can then be much reduced. If,
in addition to measuring that time-dependent asymmetry in that
channel, one also measures the rates for the isospin related
channels, one has, in principle, enough information to determine
sin(2$\alpha$). Unfortunately,  the $\pi^0\pi^0$ rate is expected
to be small, so that it may be some time before the experimental
uncertainties of this approach are small enough that the result is
actually improved by it. However even an upper bound on the
neutral pion rate can provide useful constraints \cite{bounds}.

Electroweak penguin effects can also be considered in an isospin
analysis, by writing the isospin structure of the $Z$-boson decay.
However, since this decay has isospin 1 as well as isospin 0
parts, there is a $\Delta I = 3/2, I_{\rm final}=2$ contribution,
and this cannot be separated from the tree term via any
multichannel analysis. This results in some residual theoretical
uncertainty in the extraction of $\alpha$, but it is significantly
smaller than that from the gluonic penguin contribution without
isospin analysis.

A similar situation makes isospin analysis useless in separating
tree and penguin parts for $b\rightarrow c\bar c d$ channels such
as $D^+D^-$. Here both the tree and penguin contributions are pure
$\Delta I = 1/2$, so there is no way to distinguish them via their
isospin structure.

\subsection{SU(3) Symmetry}

One can get further relationships between different processes if
one extends the idea of isospin to the full flavor SU(3), which
treats the three lightest quarks as a degenerate triplet. In
particular the subgroup of SU(3) known as U-spin under which the
down and strange quarks are a doublet gives lots of interesting
relationships between amplitudes \cite{gronauetc}. As with any
approximate method, the challenge here is to estimate the size of
possible corrections from symmetry breaking effects, that is to
estimate the theoretical uncertainty in the predictions. One can
distinguish three different types of SU(3) breaking effects. First
there are kinematic factors that occur because of the different
quark (and hence different meson) masses give different phase
space factors. These may be large but can be well-estimated and
lead to small theoretical uncertainties for any given set of
channels.  Second there are the factors of $F_\pi$ (or $f_\pi$)
versus the similar factors for the kaon.  These are measured
numbers so, where a vector or pseudoscalar meson is directly
produced by a $W$, they again lead to no significant
uncertainties.  However when the local operator that produces the
light meson is not an axial current then the corresponding ratio
is not so well determined. Calculations often use the known ratio
of $F$ (or $f$) factors to estimate the SU(3) breaking in such
cases also, but now the uncertainty is not so well-controlled.
Finally there are cases where the prediction depends also on
assuming an SU(3) relationship between the phases of decay
amplitudes. Results sensitive to this assumption may have a larger
theoretical uncertainty.

 The application of SU(3)symmetry can
allow one to use measured penguin-dominated amplitudes such as
$B\rightarrow K\pi$ to constrain the penguin contribution to a
tree-dominated amplitude such as $B\rightarrow \pi\pi$.  This
provides a collection of additional approaches to fix the CKM
parameter $\gamma$ from the combined $\pi\pi$ and $K \pi$ data
\cite{gamma}.

Another value of both Isospin and SU(3) relationships is that they
provide a window to search for effects of physics beyond the
Standard Model.  There are a number of cases where possible new
physics effects do not respect the relationships predicted by
these symmetries \cite{trojanpenguins}. Tests of these
relationships may then provide a window for new physics.

\subsection{Lattice Calculations}

Perhaps the best way to include hadronic physics and QCD effects
in a calculation of the matrix element of any operator is to use
lattice QCD methods.  Methods to treat heavy-light mesons on the
lattice have been developed and are steadily improving. There are
a number of cases where this method will eventually yield
theoretical predictions with well controlled errors. Lattice
calculation is particularly useful for quantities such as the
$B$-mixing matrix element which is a one-particle to one-particle
transition, or $f_B$, which is a one-particle to vacuum
transition.  For one particle to  multiparticle transitions (where
multi here means two or more) the problem of including final state
interactions is not solved by lattice calculations.  These
calculations are performed in Euclidean space-time and require
analytic continuation to give the actual physical result. The
uncertainties introduced by this step are difficult to quantify
and can be large.

There are basically four sources of uncertainties in lattice of
calculations of the one-particle to one-particle (or one to
zero-particle) matrix elements.  The first is the statistical
reliability of the Monte-Carlo treatment.  This is simply a matter
of doing enough calculation, and is very well understood. Second
there are the extrapolations and scale-matching to match the
finite-volume, finite-lattice-spacing  parameters and results with
the infinite-volume continuum quantities.  Again the process is
highly developed and for the most part in good control.  Third are
the methods of handling the heavy quark on the lattice, which are
also now quite well-developed.  The critical last ingredient in
this progression is for the lattice calculation to be
``unquenched''.  This means that the lattice allows the
development of virtual light quark-antiquark loops.  Such
calculations require significantly more computer time than the
corresponding ``quenched calculation'' which suppresses quark-loop
effects.  Unquenched calculations are beginning to appear, for
example for the matrix element that is relevant to the mixing
between $B$ and $\bar B$ mesons.  There then remains some
extrapolation in the light quark masses and in the number and
degeneracies of the light quarks. The prospect is that all sources
of uncertainty can be investigated, and that, at least for some of
the critical quantities, the lattice will eventually provide the
most accurate and well-controlled estimates of the matrix
elements. Well-controlled here means that the uncertainty in the
estimate can be reliably constrained.

\subsection{Quark-Hadron Duality}

Even with all these methods we are again and again confronted with
data that cannot be interpreted without further input.  We are
reduced to using models, or to making further assumptions.  One
commonly used assumption goes under the name of ``quark-hadron
duality''.  This is the assumption that if I can calculate a
quantity, such as an inclusive rate, at the quark level then that
calculation must also give the correct answer at the hadronic
level.  In a situation where we can average over a range of
energies one can indeed prove that this must be true for certain
averages, for example the energy-averaged total cross-section for
electron-positron collisions to produce hadrons.  On the other
hand it is clear that if we look in detail at any process the
quark result, calculated at low order in QCD, can not reproduce
all the details of the hadronic spectrum correctly.  In
particular, thresholds or end-points of spectra are different for
quarks and for mesons.  Perturbative quark calculations know
nothing about resonance masses, at least not in any fixed-order
calculation.

 In a $B$ decay we cannot average over energies, the energy of the
decay is set by the $B$ mass.  Even so it is popularly believed
that inclusive $B$ decays can be well-described using the
assumption of quark hadron duality. At the quark level we can
calculate the $b$-quark decay. Now we assume that gives the
inclusive meson decay correctly, because, if the quark has decayed
it must hadronize to something.  The level of assurance with which
one can make an estimate for the corrections to this approximation
varies with the process.  For inclusive semi-leptonic decays
integrating over lepton momenta provides integration over a range
of hadron invariant mass. This can be expected to reduce the
corrections. It has thus been argued that these are very small in
the inclusive semileptonic case  \cite{bigiuraltsev}.

The demands of realistic measurements can also dilute the power of
quark-hadron duality.  Consider for example inclusive
semi-leptonic decays of $B$ mesons to hadrons that contain no
charm.  In principle the measurement of this total rate can be
used to extract a value for the CKM parameter $V_{ub}$, if we can
calculate the expected rate.  We assume quark-hadron duality gives
an accurate result for the full inclusive rate, by the arguments
given above.  However in any experimental measurement, we must
make some kinematic restriction in order to exclude backgrounds
coming from the much larger rate of decays to hadrons containing
charm quarks.  This introduces dependence on details of the
spectrum, rather than just a particular integral of it.

 There is more than one way to choose the kinematic cut:  one can for
example restrict the electron momentum to be large enough that
charm production is excluded; or one can restrict the hadronic
invariant mass to be small enough to exclude charm.  Because of
the unseen neutrino these restrictions are not identical.  Each
keeps some fraction of the total rate.  To extract $V_{ub}$ we
must know what that fraction is.  But to calculate that fraction
we are looking at details of the spectrum for which the use of a
quark-level calculation may not be so safe.
 Recent work has suggested using some
combination of cuts on hadron mass and on lepton invariant mass
(which requires neutrino reconstruction). A carefully chosen
combination can minimize sensitivity to the spectrum end-point
details.  One can also make some tests as to the stability of the
result as the cut prescription is varied \cite{bauer, rothstein}.

\subsection{Models and Other Approximations}

In many other channels, even once one uses QCD-improved
factorization calculations one needs to know a meson-meson
transition matrix and/or quark distribution functions for both
initial and final state particles to calculate a rate.  Lattice
calculation, or measurement in a semi-leptonic decay, can be used
to fix the transition matrix element.  In certain cases one
obtains self-consistent quark distribution functions using
light-cone QCD arguments.  Or one can parameterize these
distributions, for example by their moments, and use some set of
measurements to fix the set of parameters that dominate an effect
(making sure that such parameters are indeed carefully and
consistently defined in both processes).

  Finally one can simply resort to making models for the
unknown quantities. One can using rigorous limits obtained from
QCD sum rules \cite{sumrules} and from the heavy quark limit to
constrain the models and reduce the number of independent inputs
needed. However this is not sufficient to remove all model
dependence of the results. There are often still large (and not
well-constrained) uncertainties that arise in this stage of the
calculation.

\subsection{Summary}

For two-body hadronic decays even QCD-improved calculations
require some  input of transition matrix elements and quark
distribution functions for the mesons in question in order to
calculate amplitudes. These input quantities can sometimes be
constrained by symmetries. Rigorous limits for some can be derived
for example from the heavy quark limit and from QCD ({\em e.g.}
the QCD sum rule methods).  Some of the quantities of interest can
eventually be accurately calculated on the lattice.  Some can be
measured in semileptonic processes. Data on a great variety of
decays will help refine our understanding. This process has
already begun.  Data from CLEO and from the two asymmetric $B$
factories gives us much to study, and will continue to do so.

Our ability to see whether different measurements yield consistent
or inconsistent values for the Standard Model parameters is only
as good as our ability to constrain the theoretical uncertainties
in a reliable fashion. As one applies any method to a multitude of
channels one can learn from experience what accuracy is obtained
and refine the method on the basis of that experience.  Because
there are indeed many possible quasi-two-body $B$ decays this
process will eventually improve our ability to constrain the
theoretical uncertainty of a given calculational method.  To
achieve this ability it is important for theorists to be as
precise and as honest as possible about the sensitivity of any
results to input assumptions or models, and to explore this
sensitivity in some detail. Only in this way can we find those
sets of measurements which truly give us sensitive tests of the
Standard Model.

\boldmath
\section{Lecture 4. Experiments to Measure $B$ Decays}
\unboldmath

In this last lecture I will review how one goes about studying
these questions experimentally.  Even though you (in this
audience) are mostly theory students, it is important that you
have some idea of how the measurements are made.  The aim of the
game is to make multiple measurements that can check Standard
Model predictions in a redundant fashion.  There are a number of
ways that physics from beyond the Standard Model could show up.
One could find inconsistent results for a particular Standard
Model parameter (or set of parameters) when determining the same
parameters by multiple independent methods.  One could find a
large $CP$-violating asymmetry in a mode for which the Standard
model predicts a small or vanishing effect.  One could find decay
modes that are predicted to be rare present at a rate different
from that expected or with a pattern of isospin or SU(3) symmetry
violations that cannot be accommodated within the theoretical
uncertainty of Standard Model predictions. Each of these
possibilities requires ongoing work on both the theory front, to
reduce theoretical uncertainties, and the experimental one, to
make all the suggested measurements.  I will focus on $B$ decay
experiments, but rare $K$-decay results also contribute to the
picture, as do the existing results on $CP$-violation in $K$
decays.

\boldmath
\subsection{Tagging $B$ Flavor}
\unboldmath

Up until now we have talked about various decays of an individual
$B$ meson as if we knew what meson we had at time $t=0$.  The
flavor conservation of strong and electromagnetic interactions
means that one produces a $b$-quark and an anti-$b$-quark in the
same event.  In general one has no {\em a priori} knowledge of
which type of neutral $B$ meson was formed at production.  One
must use other properties of the total event in order to determine
whether one had a $B^0$ or $\bar B^0$ meson at production (or at
some other known time). This process is called tagging.  For
example one can tag a $B$ meson when another $B$ meson in the same
event decays in such a way that its $b$-flavor is identifiable. An
example of a tag is a semileptonic decay; the charge of the lepton
then identifies whether it came from the weak decay of a $b$ or a
$\bar b$ quark. The tagging possibilities and efficiencies are
quite different in $e^+e^-$ collisions and in hadronic collisions,
but the requirement for tagging is common to both types of
experiments.

In principle almost every event has some tagging information.
Often this information is not precise.  For example consider the
lepton-charge tag suggested above.  If the $b$-quark decays
hadronically to a $c$-quark which then decays semileptonically
then the detected lepton comes from the decay of the $c$ instead
of that of the $b$. Assuming it came from the $b$ will give a
wrong sign tag. The spectrum of such secondary-decay leptons is
different from that of the primary ones. One can use such
additional information to improve the correctness of the tag.
However the two spectra overlap, so there will still be cases
where there is an ambiguity.  Only a probability for each tag-type
can be determined.  Each type of tag event thus has two properties
that must be understood, its efficiency, $\epsilon$, and the wrong
tag fraction, $w$ associated with it.  Some methods have very high
purity but low efficiency, others with much higher efficiency may
have lower purity.  The measure of tagging quality that eventually
determines how well we can measure a $CP$-violating asymmetry is
the product $\epsilon(1-2w)^2$.  We will see below how this comes
about.  Both the efficiency and the wrong tag fraction are
determined by a combination of Monte Carlo modelling of events and
measurements, for example from samples of doubly tagged events.  A
significant systematic uncertainty in the result for any asymmetry
arises from the uncertainty in determining the wrong tag fraction.
Since that determination is at least in part data driven, this
uncertainty will decrease as data samples increase.

\boldmath
\subsection{$e^+e^-$ Collisions}
\unboldmath

In an electron-positron collider the most efficient way to produce
$B^0$ mesons is to tune the energy to the $\Upsilon_{4s}$, since
that large resonant peak in event rate is just above threshold to
decay into either a $B^+ $ and a $B^-$ or into a $B^0$ and a $\bar
B^0$. Hence the $\Upsilon_{4s}$ decays essentially $50\%$ to each
of these states.  Furthermore, the two neutral mesons are produced
in a coherent state which, even though both particles are
oscillating as described previously, remains exactly one $B^0$ and
one $\bar B^0$ until such time as one of the particles decays. For
studies of $CP$-violation this turns out to be either a disaster
or a very useful property depending on the design of your
collider.

To observe $CP$ violation we must look for decays where one of the
two neutral $B$'s decays in a way that identifies its flavor, so
that it gives a good tag, and the other decays to the $CP$
eigenstate of interest for the study.  Then we examine the decay
rate as a function of the time, $t$, between the tagging decay
(defined to occur at $t=0$) and the $CP$-eigenstate decay.  When
the tag is a $\bar B^0$ this means that the particle which decayed
to the $CP$ eigenstate is known to have been a $B^0$ at time $t=0$
(or, for $t<0$, to be that combination which would have evolved to
be a $B^0$ at time $t=0$).  We denote this state as $B^0(t)$.  Its
decay rate as a function of time is given by
\begin{equation}
R(B^0(t)\rightarrow f) = |A(B^0\rightarrow f)|^2e^{-\Gamma |t|} [1
+(1-|\lambda_f|^2)\cos(\Delta m t) +Im\lambda_f \sin(\Delta m t)]
\end{equation}
where once again $\lambda_f= (q/p)[A(\bar B^0\rightarrow
f)/A(B^0\rightarrow f)]$. In this equation and all following
discussion of $B_d$ decays we neglect $\Delta \Gamma$, and,
equivalently, assume $|q/p|=1$.  (The corresponding formulae for
$B_s$ decays are a little more complicated as this approximation
cannot be used in that case, you can find them in the textbooks
\cite{texts}. ) Likewise, the rate when the tagging decay is a
$B^0$ is
\begin{equation} R(\bar
B^0(t)\rightarrow f) = |A(B^0\rightarrow f)|^2 e^{-\Gamma |t|}
[|\lambda_f|^2 +(|\lambda_f|^2-1)\cos(\Delta m t)
-Im\lambda_f\sin(\Delta m t)] \ .
\end{equation}

Notice that if we were to integrate over all times, $-\infty \le
0\le \infty$ the term proportional to sin$(\Delta M t)$ would
integrate to zero.  This would destroy our sensitivity to the
$CP$-violating quantity Im$\lambda_f$.  We must measure the
asymmetry between $B$ tags and $\bar B$ tags as a function of time
to avoid this cancellation.  For a symmetric electron positron
collider running at the $\Upsilon_{4s}$ this is essentially
impossible.  (This is the disaster referred to above.) The two $B$
mesons are produced with small momenta.  Even with the best
detectors one cannot accurately measure the difference in distance
from the collision point of the two decays.  Indeed the size of
the beam-beam interaction region is typically sufficient to
destroy any possibility of resolving this difference.  Hence
cannot measure the time-difference between the decays.  Pier
Oddone suggested an idea that allowed $B$ factories to be built to
tackle $CP$ violation \cite{oddone}. The idea was to build two
storage rings with different energies and collide the electrons
and positrons so that the $\Upsilon_{4s}$, and likewise the pair
of $B$'s to which it decays, are produced moving, with a
significant relativistic gamma-factor.  Then the physical
separation of the decay vertices of the two $B$'s is increased via
the time dilation of the decay half-life.  (A decay vertex is the
point from which the tracks of the particles produced in the decay
diverge.)  In this case one can indeed, using a precision tracking
device known as a vertex detector, resolve the two decay vertices
and measure their separation with a resolution that is small
compared to the average separation. Furthermore, since any
transverse motion of the $B$ mesons is small compared to the
overall center-of-mass momentum, the distance between the decays
(in the higher-energy beam direction) gives a good measure of the
time between them. The uncertainty in the production point due to
beam size is irrelevant for this measurement, as we are not
concerned with time from production, but only the time between the
two decays. Thus the initial coherent state gives a beautiful
prediction for a measurable time-dependent asymmetry.  The
experiment has many internal cross checks that can be made to
confirm that the effect is seen as predicted.  For a detailed
discussion of the physics capabilities of such a facility see for
example the BaBar Physics Book, which is available via the web
\cite{babarbook}.

To see how the tagging efficiency affects the result consider how
the measured asymmetry is related to the actual asymmetry. The
total number of events that we count as $B$-tagged events is
$\epsilon (N_B (1-w) + N_{\bar B}w) $ where $N_B$ and $N_{\bar B}$
are the actual numbers of $B$ and $\bar B$ events produced.
Likewise the total count of $\bar B$ events is $\epsilon (N_B w +
N_{\bar B}(1-w)) $. Thus the measured asymmetry is
\begin{equation}
a_{\rm meas} =(1-2w){(N_B - N_{\bar B})\over N_B + N_{\bar B})} =
(1-2w)a_{\rm true}
\end{equation}
where $a_{\rm true}$ is the true asymmetry.  In addition the total
number of events included in the result scales with $\epsilon$,
the tagging efficiency, since only tagged events can be used.
Since statistical accuracy grows like the square root of the
number of events, the accuracy of the measurement is proportional
to the square root of epsilon.  Combining these two facts gives
you an understanding of the earlier statement that the quality
measure for tagging is $\epsilon(1-2w)^2$.  This is sometimes
called the effective tagging efficiency.

Both asymmetric $B$ factory projects, one at SLAC \cite{pep2} and
the other at KEK \cite{KEKB}), have succeeded spectacularly in
building and operating a two-storage-ring facility together with a
detector and computer system capable of detecting and recording
all the relevant details of millions of $B\bar B$ events.
Interesting data from these facilities is now beginning to be
reported and will continue over the next several years to yield
new insights.  See the websites of the BaBar \cite{babar} and
Belle \cite{belle} experiments for details.

In addition to measuring $CP$-violating asymmetries these
facilities are also compiling and analyzing large data samples for
a variety of $B_d$ decays.  Together with measurements from the
symmetric $B$ factory at Cornell \cite{cesr}
 and its detector CLEO \cite{cleo}, this data will
considerably refine our ability to measure the $CP$-conserving
parameters and to test theoretical calculations.  I have talked in
previous lectures about the uncertainties that plague many
theoretical calculation methods, and in particular about the
difficulty in quantifying these uncertainties.  As data on
multiple modes accumulates we can refine our understanding of the
accuracy of various approaches by comparison with this data.

\subsection{Proton Colliders}

Because the $B$-factory machine's are optimized to run at the
$\Upsilon_{4s}$ they are below the threshold to produce any $B_s$
mesons.  In principle they could do so by running at the
$\Upsilon_{5s}$.  The smaller peak height of this resonance,
together with the fact that it has many possible decay channels
combine to make the production rate for $B_s \bar B_s$ pairs
significantly lower than that for $B_d$ at the $\Upsilon_{4s}$.
The machines would have to be be re-optimized to run at this
higher energy, which itself is not a simple change.  All these
factors combine to make it unlikely that this will be attempted
any time soon, while there is still so much to learn about the
$B_d$ decays. So for measurements of $B_s$ decays, and also for
those of baryons containing $b$-quarks, we need to look elsewhere,
to hadron colliders.  For the time being that means the Fermilab
TeVatron \cite{tevatron}, eventually it will also mean LHC
\cite{lhc}  at CERN.

At a hadron collider the $b$ and $\bar b$ quarks hadronize
independently and each $B$ meson is part of a large jet of many
particles.  Many more $B$'s are produced in high energy
hadron-hadron collisions than in an electron-positron $B$ factory.
Hadronic collisions also produce many other types of events, with
yet higher cross-sections.  Thus, for these experiments, it is
critical to devise ways to identify $B$-events fast enough to
trigger the system to record the event.  The trigger is typically
two charged tracks emerging from a $B$-decay vertex that is
separated from the beam-beam collision region.  The design of the
trigger and its efficiency is a very important and challenging
feature of these experiments.  The triggering requirements
restrict the decay channels that can be studied in a hadronic
environment.  The methods and efficiencies for tagging the flavor
of the produced $B$ are also quite different in the hadronic case
than in the electron-positron $B$ factory environment.  The
tagging particle may be a charged $B$ or a baryon, or it may be
deduced from properties of the leading particles in the jet
containing the neutral $B$.  Furthermore, since the two $b$-quark
(or antiquark) containing particles are not in a coherent state,
the time evolution of the $CP$-study particle (and  also the
tagging particle if it is a neutral $B$-meson)  starts at
production time. There are a number of interesting quantities that
can only be studied in a hadron facility, others where the two
types of machines are competitive, and some where the
electron-positron machines have unique capabilities. Both
approaches are needed to gather all the information we would like
to have.

An example of a quantity where hadron collider results will be
important is the determination of the side $V_{td}$ of the
unitarity triangle. Currently this quantity is determined by
measuring the $B_d$ mass difference.  However there is a
significant theoretical uncertainty that arises when relating the
measurement to the parameter $V_{td}$.  Much of this uncertainty
would be removed by a measurement of the $B_s$ mass difference as
well as that for $B_d$.  The ratio of the two mass differences
gives $V_{td}/V_{ts}$ with relatively controlled theoretical
uncertainties.  If the value predicted by the Standard Model is
correct this measurement can be done at Fermilab in the CDF
experiment, probably within the next couple of years.

There has been a detailed study of the opportunities for $B$
physics in Run II at Fermilab \cite{TevatronB}. The CDF \cite{cdf}
and D-Zero \cite{D0} detectors have just completed upgrades and
are beginning to take data, including some $B$-physics-triggered
data. In addition a new experiment,known as BTeV, with a detector
optimized for $B$-physics capability, is planned \cite{btev}.  At
CERN there is also such an experiment planned, known as LHCB
\cite{lhcb}. These detectors will give expanded $B$ physics
capability and perhaps allow some rare modes to be studied, with
branching fractions that are too small to measure in the current
experiments.  (After my talk I was told there is also a study
underway of a possible future $B$ experiment at HERA, a follow-up
to the HERA-$B$ experiment \cite{herab} using a wire target in the
proton beam of that $e$-$p$ collider.) Another future option is an
intense $Z$-production facility at a linear collider, where study
of $Z\rightarrow b \bar b$ decays can yield useful additional
possibilities.)  All in all, the problem has many aspects.  The
complementarity of the different experiments will allow a rich
program of measurements.  Eventually we will have a clear picture
of whether the pattern of results matches the Standard Model or
requires some physics beyond the Standard Model to describe the
data.

\subsection{Some Final Remarks}

As theorists search for ways to extract interesting information
from $B$ decays they will often describe desired measurements that
are beyond present capabilities.  This is not new.  When Bigi and
Sanda \cite{bigisanda} first talked about $CP$-violation in $B$
decays we did not know the $B$ lifetime, so the measurements that
they proposed seemed out of reach.  Sometimes nature is kind and
the numbers work out better than present knowledge suggests.
Sometimes clever technical ideas, such as the asymmetric $e^+e^-$
collider, extend our experimental reach.  Improvements in the
technology of particle tracking and particle identification have
been essential in the $B$ factory experiments and will continue to
be so for BTeV and LHCB.  The history of discovery in science
continues because measurements deemed impossible in one era become
feasible with new developments.  Likewise new developments on the
theory side, such as new techniques for unquenched lattice
calculations are important, as they allow more measurements to be
interpreted with good control of theoretical uncertainties.

To conclude this lecture series I would like to remind you that
the aim of the game in studying $CP$ is to examine this
least-explored corner of the Standard Model in two ways.  The
first is to pin down the value of the remaining Standard Model
parameters.  The second is to test whether multiple measurements
give consistent answers, both for the parameters and for other
Standard Model predictions.  The hope is that any discrepancy will
be a clue to the nature of physics beyond the Standard Model,
physics that can, for example, change the relative phase of a
mixing amplitude compared to a decay amplitude.  Indirect searches
for new physics, such as these $B$ physics probes, are a blunt
instrument. Many extensions of the Standard Model may predict
similar effects, for example additional contributions to the
mixing.  The challenge to theorists is to reduce theoretical
uncertainties to the point that we sharpen that instrument enough
to see the effects if they are there, rather than losing them in
the ranges of possible answers given by our poor control of
hadronic physics effects.  This work is well begun, but there is
more to do.  I hope some of the students here will make
interesting contributions to it in the near future.


\begin {thebibliography}{99}
\bibitem{TevatronB}
Talks and a preliminary draft of the report can be found
at\hfill\break www-theory.fnal.gov/people/ligeti/Brun2/
\end{thebibliography}
\end{document}